\documentclass[a4paper,12pt]{scrartcl}
\pdfoutput=1

\usepackage{amsmath}
\usepackage{commath}
\usepackage{amsfonts}
\usepackage{amssymb}

\usepackage[caption=false]{subfig}
\usepackage{floatrow}

\usepackage[pdftex]{graphicx,color}
\usepackage{siunitx}
\usepackage{textcomp}

\newcommand{\be}{\begin{equation}}
\newcommand{\ee}{\end{equation}}
\newcommand{\beq}{\begin{eqnarray}}
\newcommand{\eeq}{\end{eqnarray}}

\usepackage{hyperref}

\usepackage{authblk}

\begin{document}


\title{Free energy function of dislocation densities by large scale atomistic simulation}


\author[1]{Christoph Begau}
\author[1,2]{Godehard Sutmann}
\author[1]{Alexander Hartmaier}
\affil[1]{Interdisciplinary Centre for Advanced Materials Simulation (ICAMS), Ruhr-University Bochum, D-44801 Bochum, Germany}
\affil[1]{J\"ulich Supercomputing Centre (JSC), Forschungszentrum J\"ulich (JSC), D-52425 J\"ulich, Germany}

\date{}
\maketitle



\begin{abstract}
This paper discusses the free energy of complex dislocation microstructures, which is a fundamental property of continuum plasticity.
In the past, multiple models of the self energy of dislocations have been proposed in the literature that partially contradict each other.
In order to gain insight into the relationship between dislocation microstructures and the free energy associated with them, instead of deriving a model based on theoretical or phenomenological arguments, here, these quantities are directly measured using large scale molecular dynamics simulations.
Plasticity is induced using nanoindentation that creates an inhomogeneous distribution of dislocations as the result of dislocation nucleation and multiplication caused by the local deformation.
Using this approach, the measurements of dislocation densities and free energies are \textit{ab-initio}, because only the interatomic potential is defining the reaction of the system to the applied deformation.
The simulation results support strongly a linear relation between the scalar dislocation density and the free energy, which can be related very well to the classical model of mechanical energy of straight dislocations, even for the complex dislocation networks considered here.
\end{abstract}


\section{Introduction}
Dislocations are the main carriers of plastic deformation and are thus of high importance for crystal plasticity simulations.
Crystalline materials store energy in the elastic fields of dislocations and thus a description of the relationship between the dislocation microstructure and the free energy associated with them is relevant for such simulations as the partial derivatives of the energy are related to stress and the evolution equation of the system.

For this reason several models describing a free energy density function for dislocation densities, based on different assumptions, have been suggested by various authors.
In the mechanical literature \cite{Steinmann1996, Svendsen2010,Forest2013} a general form of this energy function that depends on three constitutive variables is introduced as
\be
\psi = \psi(\boldsymbol{\alpha}, \rho, \epsilon)\quad,
\label{Eq:energyDensPsi}
\ee
where $\boldsymbol{\alpha}$ denotes the dislocation density tensor, originally introduced by Nye \cite{Nye1953}, Kr{\"o}ner \cite{Kroner1958} and Ashby \cite{Ashby1970}, as a characteristic quantity of the effective plastic deformation of a crystal, which is immediately related to the geometrically necessary dislocation (GND) density.
In contrast to this geometrically motivated density, the scalar dislocation density $\rho$ describes the total length of all dislocations within a volume and is equal to the sum of the GND density and the density of statistically stored dislocations (SSD). 
The net Burgers vector content in a volume containing only SSDs is zero.
Furthermore, $\epsilon$ denotes an additional elastic deformation that can be present if an external force acts on the body.

As noted before, different forms of a free energy function have been suggested, however, they are at least partially contradicting each other.
Steinmann and coworkers \cite{Steinmann1996,Menzel2000} suggest free energy functions that are based on a quadratic ansatz of the GND density to compute a norm of the dislocation density tensor.
Other authors suggest energy functions that are based on the GND density of individual slip systems, e.g.\ the model of Gurtin et al.~\cite{Gurtin2007} assumes that the free energy is dependent on the sum of the squared screw and edge GND density on multiple slip systems.
Similarly, Ohno et al.~\cite{Ohno2007} proposed a model that is based on the same densities, but does not feature a quadratic form and instead, the self-energy of dislocations is related to the square-root of the sum of squared screw and edge GND densities instead.
These models all have in common that they only account for the GND densities.
Groma et al.~\cite{Groma2007} suggest an energy function that includes both a quadratic term of the GND density and a contribution of the total dislocation density in the form $\rho \log \rho$.
In contrast to all these models, Berdichevsky~\cite{Berdichevsky2006} suggests a function that does not depend on GNDs at all and scales linearly with the total dislocation density as long as this density is significantly smaller than a saturation density.
If this saturation value is exceeded, deformation cannot be accommodated anymore by dislocations and other mechanisms, e.g.\ the formation of subgrains, will be activated.
A more detailed overview and comparison of different free energy functions in crystal plasticity is given e.g.\ in Refs.~\cite{Svendsen2010, Forest2013, Mayeur2014}.

While the form of the free energy function of dislocation networks is subjected to different approaches, the energy of individual straight edge and screw dislocations is well understood.
The associated energy of such dislocations can be split into two components -- energy stored in the dislocation core and elastic energy stored in the volume of the crystal.
The elastic energy of a straight dislocation segment is estimated as \cite{HullBacon}
\be
E_{\mathrm{el}}^{\mathrm{dis}}=A \cdot \frac{G \cdot l b^2}{4\pi}\cdot \ln{\left( \frac{R}{r_0} \right)}\quad, 
\label{Eq:ElasticEnergy}
\ee
where $l$ denotes the length of the dislocation segment, $b$ the norm of the Burgers vector, $R$ a characteristic length for the size of the strain field, e.g.\ the grain size, $r_0$ the dislocation core radius, $G$ the shear modulus and $A$ a material dependent parameter, related to the Poisson's ratio $\nu$ that differs for screw ($A=1$) and edge dislocation ($A=(1-\nu)^{-1}$).
The energy of dislocation cores is more difficult to be estimated and is frequently computed by first principle and MD simulation, e.g.\ for Fe \cite{Clouet2009} or Mo \cite{Ju2004}.
In any case, the core energy is observed to scale linearly with the length of the dislocation and is on the order of a few \si{\electronvolt\per\nano\meter}, corresponding to some fraction of the elastic energy, with the exact value depending on the material.

As outlined, the correct form of an energy function based on dislocation densities is being discussed intensively in the literature, especially since experimental references are missing on this topic.
In this work, we will address this issue in order to shed light on the proper form of an energy function of dislocation densities using large scale Molecular Dynamics (MD) simulations.
Pristine defect free single crystals are deformed using nanoindentation, causing the formation of complex dislocation networks.
Most importantly, these simulations do not require assumptions on the evolution of dislocation microstructures or other aspects of crystal plasticity, but instead the plastic deformation is solely a result of the interatomic interactions.
It is noted that the nanoindentation produces GNDs, as well as SSDs, such that their respective contribution to the energy of the dislocation microstructure can be separated.
Recently, several advanced analysis methods have been developed that now permit the identification of dislocation networks including Burgers vectors within atomistic simulations \cite{Begau2012, Stukowski2012, Wang2013} and thus, dislocation densities can be precisely measured in such simulations.
Furthermore, the energy $\psi$ stored in the crystal after deformation is immediately available from the MD simulation results as well.
The defect energy of a dislocation microstructure is frequently directly related to the free energy of dislocation densities in the literature.
E.g., Gurtin et al.\ \cite{Gurtin2007} consider the defect energy, besides the classical elastic strain energy, as a part of the free energy, while similar definitions are given as well in \cite{Svendsen2002,Svendsen2010}.
Since, we measure the defect energy in equilibrated samples after deformation without external forces acting on them, the contribution of elastic strain energy can be omitted, and thus, we refer to the mechanical energy stored in the dislocation microstructure as the free energy in the following.
To the authors best knowledge, the relation of the free energy $\psi$ to dislocation densities $\rho$ and $\boldsymbol{\alpha}$ of self-assembling dislocation networks has not been studied on an atomistic scale before.

\section{Method}
\label{sec:Method}
\paragraph{Dislocation densities}
\label{sec:disDens}
For a given dislocation network, the corresponding dislocation densities can be computed straightforwardly.
The dislocation density $\rho$ is equal to the total length of all dislocations in a volume $V$
\be
\rho = V^{-1}\int_{\perp \in V} d\vec{l}_{\perp} \quad.
\label{Eq:rho}
\ee
The symbol $\perp$ denotes a dislocation, here without a distinction between edge and screw dislocations, and $\vec{l}_{\perp}$ its line tangent vector.
The GND tensor $\boldsymbol{\alpha}$, which takes into account the Burgers vector $\vec{b}_{\perp}$ of a dislocation, is defined by
\be
\alpha = V^{-1}\int_{\perp \in V}\vec{b}_{\perp}\otimes d\vec{l}_{\perp} \quad .
\label{Eq:GND}
\ee
Since $\rho$ is a scalar value and $\boldsymbol{\alpha}$ a second order tensor, it is not possible to compare these two quantities directly.
Therefore, a suitable norm is required to derive a scalar GND density $\rho_{\mathrm{GND}}$,
\be
\rho_{\mathrm{GND}} = \norm{\boldsymbol{\alpha}} \quad .
\label{Eq:rho_GND}
\ee
Different forms of the norm can be defined, e.g.\ in form of the double dot product $\boldsymbol{\alpha}\colon\boldsymbol{\alpha}$ \cite{Menzel2000}, which has the advantage of being independent of the crystallographic slip system.
However, in the present work, a different approach is followed, taking the slip system of face centered cubic (FCC) crystals into account, since this provides a more accurate description of the GND density.
The scalar dislocation density $\norm{\boldsymbol{\alpha}}$ is here derived using the method of Arsenlis~\&~Parks~\cite{Arsenlis1999}, which identifies the minimum set of dislocations in a given slip system that produces an equivalent dislocation density tensor compared to $\boldsymbol{\alpha}$.
This method provides a unique description of the minimal and therefore geometrically necessary amount of dislocations with respect to the chosen crystallographic basis.
We implemented the $L^1$ minimization scheme for a set of 18 different dislocations in the face centered cubic (FCC) slip system of type $1/2\langle110\rangle$ and $1/6\langle112\rangle$ as described in \cite{Arsenlis1999} in MATLAB using the constrained minimization routines with sequential quadratic programming.

The volume fraction occupied by dislocation cores is another possibility to define a dislocation density in a volume $V$.
Assuming that dislocation networks are composed of cylindrically shaped segments, this core density $\rho_c$ can be defined via the core volume.
\be
\lambda = \pi \int_{\perp \in V} \abs{\vec{b}_{\perp}}^2  \cdot d\vec{l}_{\perp}
\label{Eq:lb2}
\ee
\be
\rho_c = c^2 \cdot \lambda \cdot V^{-1}
\label{Eq:rho_c}
\ee
Since in certain materials the core radius $r_0$ is observed to be slightly larger than the Burgers vector norm $\abs{\vec{b}_{\perp}}$ \cite{Clouet2009} a dimensionless material dependent factor $c$ is included in the formula where
\be
r_0 = c \abs{\vec{b}_{\perp}} \quad.
\ee
The exact value of $r_0$ is not known for the simulations described in this work, and thus, the core density $\rho_c$ is estimated instead under the assumption that the volume of an atom in a dislocation core is approximately the same as in the bulk material.
With the number of atoms in dislocation cores $N(a\in\perp)$ and the total number of atoms $N$ in the volume $V$, $\rho_c$ is then defined as
\be
\rho_c = \frac{N(a\in\perp)}{N}\cdot V^{-1}\quad.
\label{Eq:rho_c_by_atoms}
\ee
The approach to decide if an atom is part of a dislocation core is discussed later.

\paragraph{Simulation setup}
In order to measure the energy of complex dislocation networks, a series of large scale MD simulations of nanoindentation in the (111) free surface of single crystalline aluminum samples has been performed.
Two different indentation depths are simulated in five cubically shaped samples with lateral sizes varying between \SI{50}{\nano\meter} and \SI{150}{\nano\meter}.
Detailed information regarding the simulation dimensions are given in Table~\ref{table:Configuration}.
In all simulations the ratio of sample size $s$ to the spherical intender tip radius $r$ is kept constant at 1:6.25, and the two maximum indentation depths $d$ depend on the indenter radius as $d=0.3r$ and $d=0.5r$.
According to the Nix~\&~Gao model \cite{NG97, Swadener2002}, $\rho_{\mathrm{GND}}$ in a hemispherical domain underneath a spherical indenter tip only depends on the radius and not on the indentation depth.
However, it has been observed in simulations \cite{Begau2012, Hua2010} that the total dislocation density $\rho$ is increasing with the indentation depth in the same volume, while $\rho_{\mathrm{GND}}$ is indeed constant.
Thus by using different indenter radii and indentation depths, different ratio of $\rho_{\mathrm{GND}}$ to $\rho$ are to be expected and possible size effects are to be identified from the results of multiple sample sizes. 
 
\begin{table}
\centering
\begin{tabular}{c|c c c c c}
Indenter radius $r$ & \SI{8}{\nano\meter} & \SI{12}{\nano\meter} & \SI{16}{\nano\meter} & \SI{20}{\nano\meter} & \SI{24}{\nano\meter}\\
\hline
Simulation box size $s$ & \SI{50}{\nano\meter} & \SI{75}{\nano\meter} & \SI{100}{\nano\meter} & \SI{125}{\nano\meter} & \SI{150}{\nano\meter}\\
Number of atoms $N$ (million) & 7.4 & 25.5 & 60.3 & 117.3 & 202.9 \\
Indentation depth $d=0.5r$ & \SI{4}{\nano\meter} & \SI{6}{\nano\meter} & \SI{8}{\nano\meter} & \SI{10}{\nano\meter} & \SI{12}{\nano\meter}\\
Indentation depth $d=0.3r$ & \SI{2.4}{\nano\meter} & \SI{3.6}{\nano\meter} & \SI{4.8}{\nano\meter} & \SI{6}{\nano\meter} & \SI{7.2}{\nano\meter}\\
Relaxation time at $d=0.5r$ & \SI{120}{\pico\second} & \SI{140}{\pico\second} & \SI{180}{\pico\second} & \SI{160}{\pico\second} & \SI{180}{\pico\second}\\
Relaxation time at $d=0.3r$ & \SI{80}{\pico\second}   & \SI{120}{\pico\second} & \SI{120}{\pico\second} & \SI{160}{\pico\second} & \SI{140}{\pico\second}\\
\end{tabular}
\caption{Variable parameters used for multiple simulations of nanoindentation.}
\label{table:Configuration}
\end{table}

To describe the interatomic interaction of aluminum, the Embedded Atom Method potential by Farkas et al.~\cite{PhysRevB.59.3393} has been used, which is particularly fitted to predict the material's elastic properties and stacking fault energies accurately.
Aluminum was chosen as a model material, since well established analysis methods exist for FCC structures and the high stacking fault energy suppresses the splitting of dislocations into Shockley partials separated by large stacking faults and deformation twinning.
Compact and less dissociated core structures are favorable since an analysis of dislocation networks is more prone to artifacts if complex reaction products of partial dislocations are present.

All simulations were performed using the massively parallel open source MD code IMD \cite{IMD,IMD2} with a simulation time step of \SI{1}{\femto\second}.
The spherical indenter tips were modeled by a purely repulsive potential \cite{PhysRevB.67.104105}, with an indentation velocity of \SI{20}{\meter\per\second} which is equal in all simulations.
Due to the different simulation sizes, between 200,000 and 600,000 MD time steps were necessary to reach the maximum indentation depth of $d=0.5r$.
After the targeted indentation depth has been reached, the indenter was hold at this position for \SI{20}{\pico\second} or 20,000 time steps, before the sample was unloaded by retracting the tip again with \SI{20}{\meter\per\second}.
During this process elastic, and to some degree inelastic deformation in the form of dislocation annihilation and absorption at the free surface, is recovered and only the residual plastic deformation remains.
The simulations were stopped once contact between free surface and indenter was lost completely for a period of at least \SI{40}{\pico\second} and the average potential energy per atom was fluctuating on the order of \SI{d-6}{\electronvolt} within an interval of \SI{20}{\pico\second}, indicating that the sample has reached a stationary state.
Since the time intervals to reach this state were observed to vary significantly, the relaxation times of each configuration are given in Table~\ref{table:Configuration}.
To save computational resources, the loading scenario of the case $d=0.3r$ is reused from the one of $d=0.5r$.
Once the indentation depth of $d=0.3r$ is reached, the current atomic configuration has been saved completely and unloading has been simulated separately using this configuration.
All simulations are performed at a constant temperature of $T=\SI{30}{\kelvin}$ controlled by a Nos\'{e}-Hoover thermostat in initially stress and defect free samples.
Periodic boundary conditions are applied in the lateral directions of the sample, an open boundary at the indented surface on the top and several atomic layers at the bottom are fixed to avoid rigid body motion during indentation.

In order to measure the energy of the dislocation microstructure in the deformed samples, a reference value of the potential energy per atom $E_{\mathrm{pot}}^{\mathrm{ref}}$ in a stress and defect free single crystal at the finite temperature of $T=\SI{30}{\kelvin}$ is determined. 
For this purpose, an aluminum cubic single crystal of size \SI{10}{\nano\meter} is created and equilibrated using an NPT ensemble with zero external pressure in a fully periodic simulation setup.
Once the system is equilibrated, the average potential energy is computed over all atoms and several thousand time steps, yielding a reference value of $E_{\mathrm{pot}}^{\mathrm{ref}}=\SI{-3.356116}{\electronvolt}$.

\paragraph{Analysis}
Suitable data analysis of the atomistic system is crucial in this study, because it is necessary to characterize the microstructure in high detail to distinguish the influence of different lattice defects on the energy of the microstructure in the system.
Primarily, the focus is placed on the extracted dislocation network, however, other crystal defects including vacancies, intrinsic and extrinsic stacking faults, crystal surfaces and -- as discussed later -- to a small degree even twin- and grain-boundaries, need to be identified as well to relate the total energy stored in the deformed crystal reliably to the dislocation densities.
To identify dislocation networks directly from atomistic data, an analysis approach based on the Nye-tensor method and geometrical identification of dislocation lines and stacking faults is used \cite{Begau2011,Begau2012}.
This method is available in the IMD code for an on-the-fly analysis and permits the creation of dislocation networks in high temporal resolution even for a very large number of atoms.
The derived representation of dislocation networks consists of a set of connected curves in space, each one associated with a Burgers vector.
Furthermore, the relationship of dislocations and atoms is preserved, thus it is possible to map from the continuum description of the dislocation networks to the atomistic data and vice versa.
Using this geometrical representation of the dislocation network, the dislocation densities $\rho$ (Eq.~\ref{Eq:rho}), $\boldsymbol{\alpha}$ (Eq.~\ref{Eq:GND}) and $\rho_c$ (Eq.~\ref{Eq:rho_c_by_atoms}) can be computed precisely for any arbitrarily shaped volume.

For an initial classification of defects, the Bond-Angle analysis scheme, originally published by Ackland \& Jones \cite{Ackland2006}, is used in a simplified version as described in \cite{Begau2011}.
As noted before, aluminum was chosen because of its high stacking fault energy that should impede deformation twinning.
Nonetheless, as discussed later in detail, twinning and subgrain formation are still observed close to the indented surface in some simulations.
Since the focus is placed on the energy densities of dislocation networks, these grain-boundaries are identified in a post-processing step that constructs a surface mesh as a representation of grain- or phase boundaries as described in \cite{Begau2014}.
In order to ignore the energy contribution in these defects, any atom within a distance of \SI{1.2}{\nano\meter} (about three times the lattice constant) is marked as being biased by a grain-boundary and is not considered in the subsequent analysis steps.
Furthermore, the free surfaces are detected as well by identifying all atoms having less than ten nearest neighbors.
Atoms that are neighbored to three or more surface atoms are considered as part of the surface as well.

In total, each atom is assigned to one of six classes:
\begin{itemize}
 \item Bulk: an atom at an FCC lattice site
 \item Stacking fault: an atom at a hexagonal closed packed (HCP) lattice site
 \item Dislocation core: an atom that is identified as part of a dislocation core using the Nye-tensor method as described in \cite{Begau2012}. The norm of the numerically computed resultant Burgers vectors of these atoms exceed \SI{0.07}{\nano\meter} and are thus on the order of the shortest possible stair rod Burgers vector $1/6\langle110\rangle$, which is having a length of \SI{0.095}{\nano\meter}.
 \item Grain-boundary: an atom near a grain-boundary
 \item Surface: an atom at a free surface
 \item Other defects: an atom that does not fit in any of these classes. Most commonly these atoms are located at vacancies.
\end{itemize}
For the analysis of energies, the current potential energy of the system is compared to the reference energy to compute the difference $\Delta E$ per atom
\be
\Delta E = E_{\mathrm{pot}} - E_{\mathrm{pot}}^{\mathrm{ref}} \quad .
\ee
Thus, the change in the mechanical energy between two configurations in equilibrium at a constant temperate is measured.
Because elastic deformations are recovered during unloading and relaxation, this defect energy is considered here as the free energy of dislocation networks, following the mechanical literature as noted before.
The simulations are performed at finite, but constant temperature, and therefore, the value of $\Delta E$ of a single atom is affected by thermal noise and only the average of a sufficiently large number of atoms is significant.
In this case, the thermal noise is strongly canceled out and can be neglected.
Even in the smallest sized simulation ($r=\SI{8}{\nano\meter}$ and $d=0.3r$) about 2800 atoms are located in dislocation cores, whereas in larger cases this value is on the order of $10^4$ or even $10^5$.
Due to this large number of atoms, the effect of temperature fluctuations is small and the energy stored in different types of defects can be measured accurately without prior quenching at $T=\SI{0}{\kelvin}$ as it is required for small sized systems.

The volume $V$ of the deformed samples after excluding subgrains and free surfaces has been measured by the construction of a surface mesh to account for these entities and as well for the indenter imprint and pileup.
However, the change in volume was found to be less than 2\% with respect to the initial cubic volume, so all quantities are related to the initial system volume in the following.


\section{Results and discussion}
\label{Sec:Results}

\subsection{Analysis of the full sized samples}
\begin{figure}[tbp]
	\centering
		\sidesubfloat[]{\includegraphics[width=0.90\textwidth]{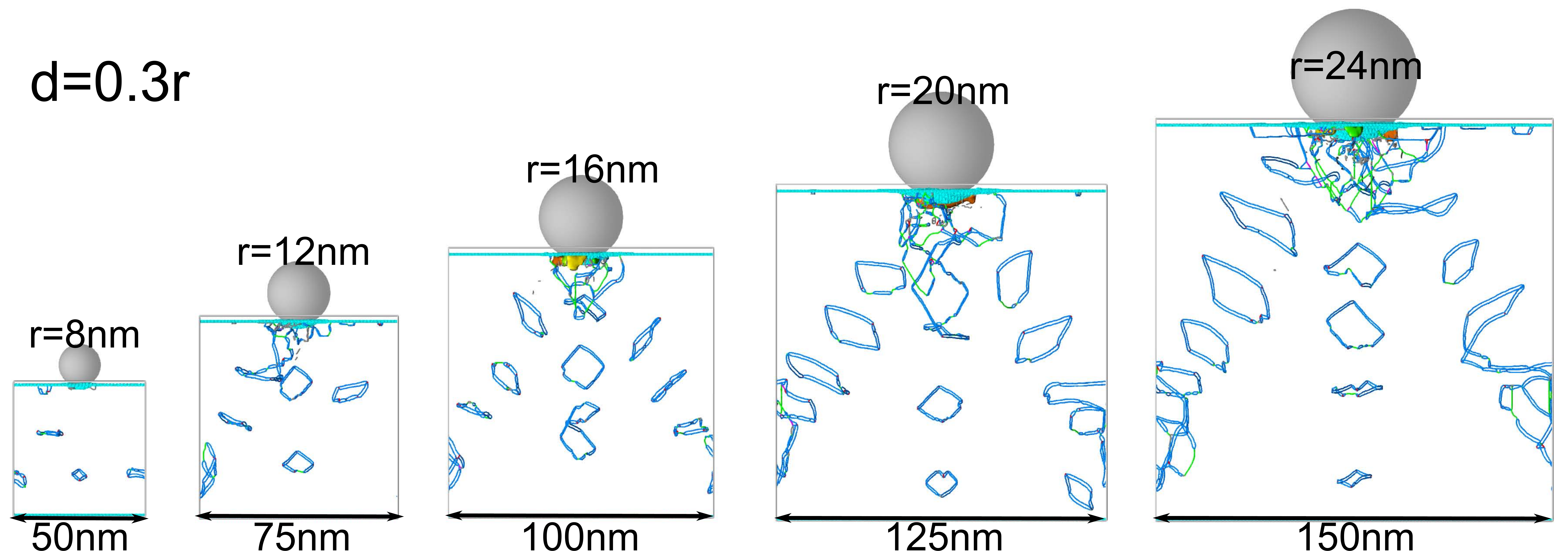}}\\
		\sidesubfloat[]{\includegraphics[width=0.90\textwidth]{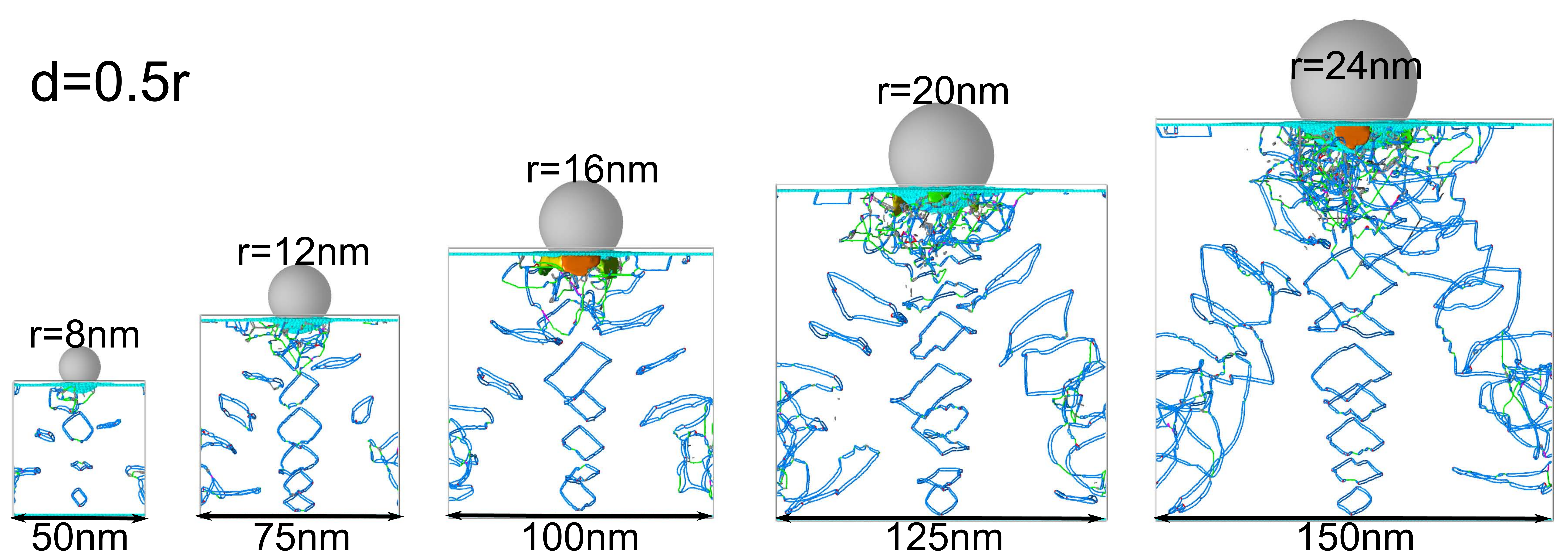}}
	\caption{The residual dislocation networks after unloading and equilibration. The indenter tips, here shown as gray spheres, have lost contact with the free surface.}
	\label{fig:DisloationNetworks}
\end{figure}

\begin{figure}[tbp]
	\centering
		\includegraphics[width=0.45\textwidth]{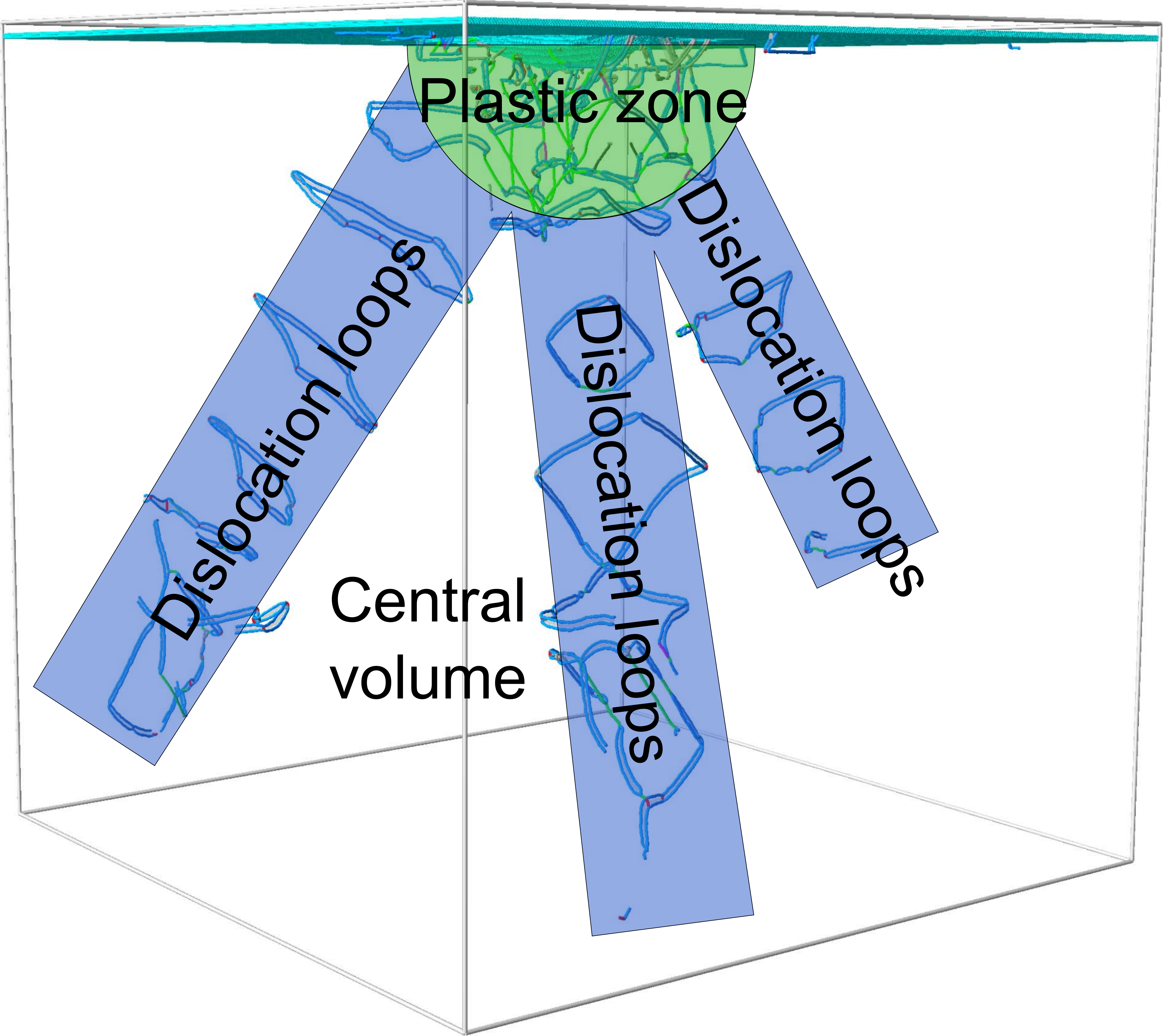}
	\caption{Illustration of the inhomogeneous distribution of dislocations in the simulation with $r=\SI{24}{\nano\meter}$ and $d=0.3r$. }
	\label{fig:InhomoDist}
\end{figure}

The residual dislocation networks of all ten unloaded and equilibrated samples are shown in Fig.~\ref{fig:DisloationNetworks}.
In this figures, far higher dislocation densities are visible for the cases with an indentation depth of $d=0.5r$ compared to their counterparts of $d=0.3r$, which is caused by the larger induced strains.
\floatsetup[figure]{subcapbesideposition=top}
However, with the exception of the smallest samples, a similar inhomogeneous distribution of dislocations is visible and three different domains can be identified.
Very high dislocation densities are present in an approximately semi-spherical region underneath the indenter imprint, often termed the plastic zone \cite{NG97}, whereas the central region of the samples is completely free of dislocations.
Further dislocations are located in a third domain that consists of prismatic dislocation loops gliding along \textlangle110\textrangle-directions.
A detailed example of this configuration is given in Fig.~\ref{fig:InhomoDist}.
Such a configuration is commonly reported for indentations in the (111)-plane on FCC material using MD \cite{Lee2005} and discrete dislocation dynamics simulations \cite{Chang2010}.
It is noted here that, due to periodic boundary conditions, gliding prismatic dislocation loops on two of the three directions collide with periodic images and form complex reaction products.

\floatsetup[figure]{style=plain,subcapbesideposition=top}
\begin{figure}[tbp]
	\centering
		\begin{minipage}{0.44\textwidth} 
		 \sidesubfloat[]{\label{fig:GrainsUnderIndenterA}\includegraphics[width=0.98\textwidth]{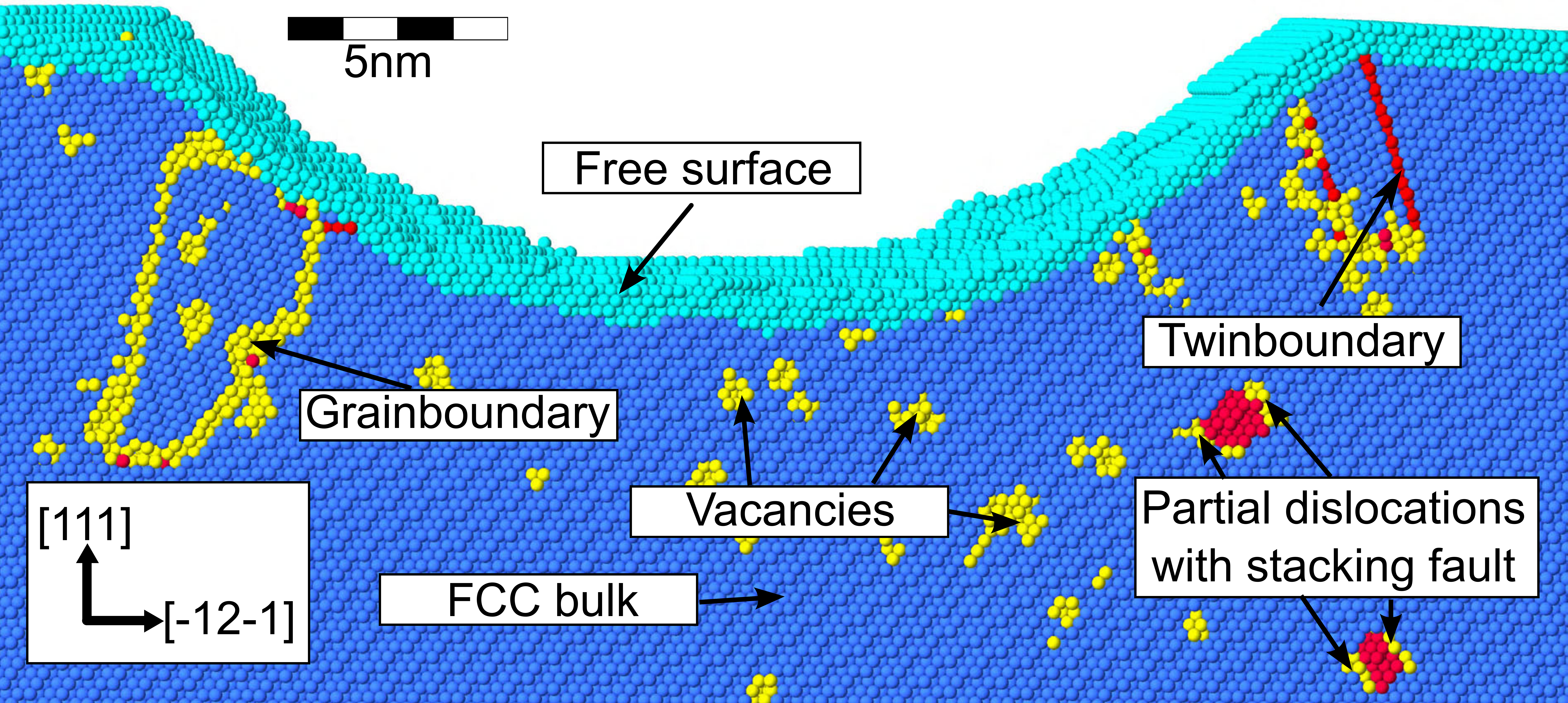}}\\
		 \sidesubfloat[]{\label{fig:GrainsUnderIndenterB}\includegraphics[width=0.98\textwidth]{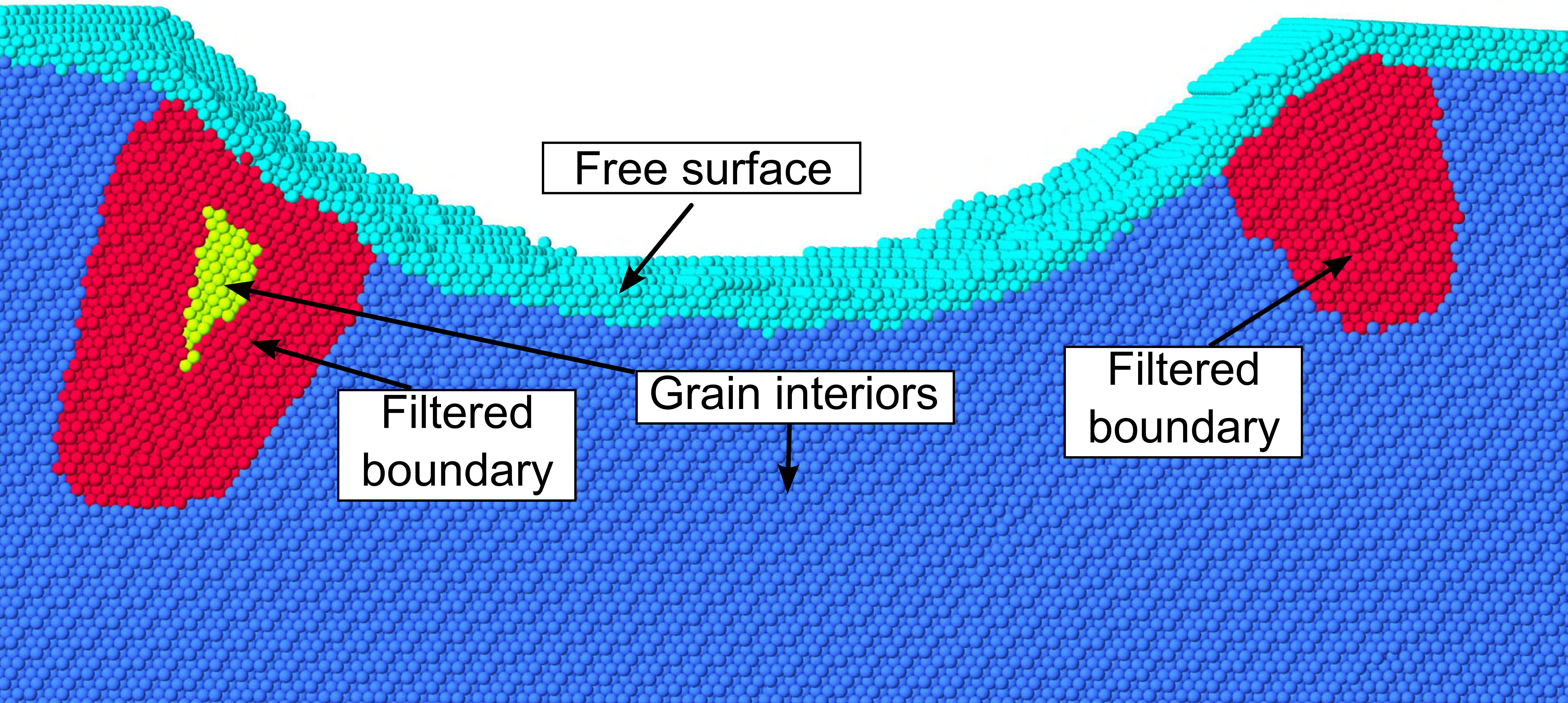}} 
		\end{minipage}\qquad
		\begin{minipage}{0.44\textwidth} 
		 \sidesubfloat[]{\label{fig:GrainsUnderIndenterC}\includegraphics[width=0.98\textwidth]{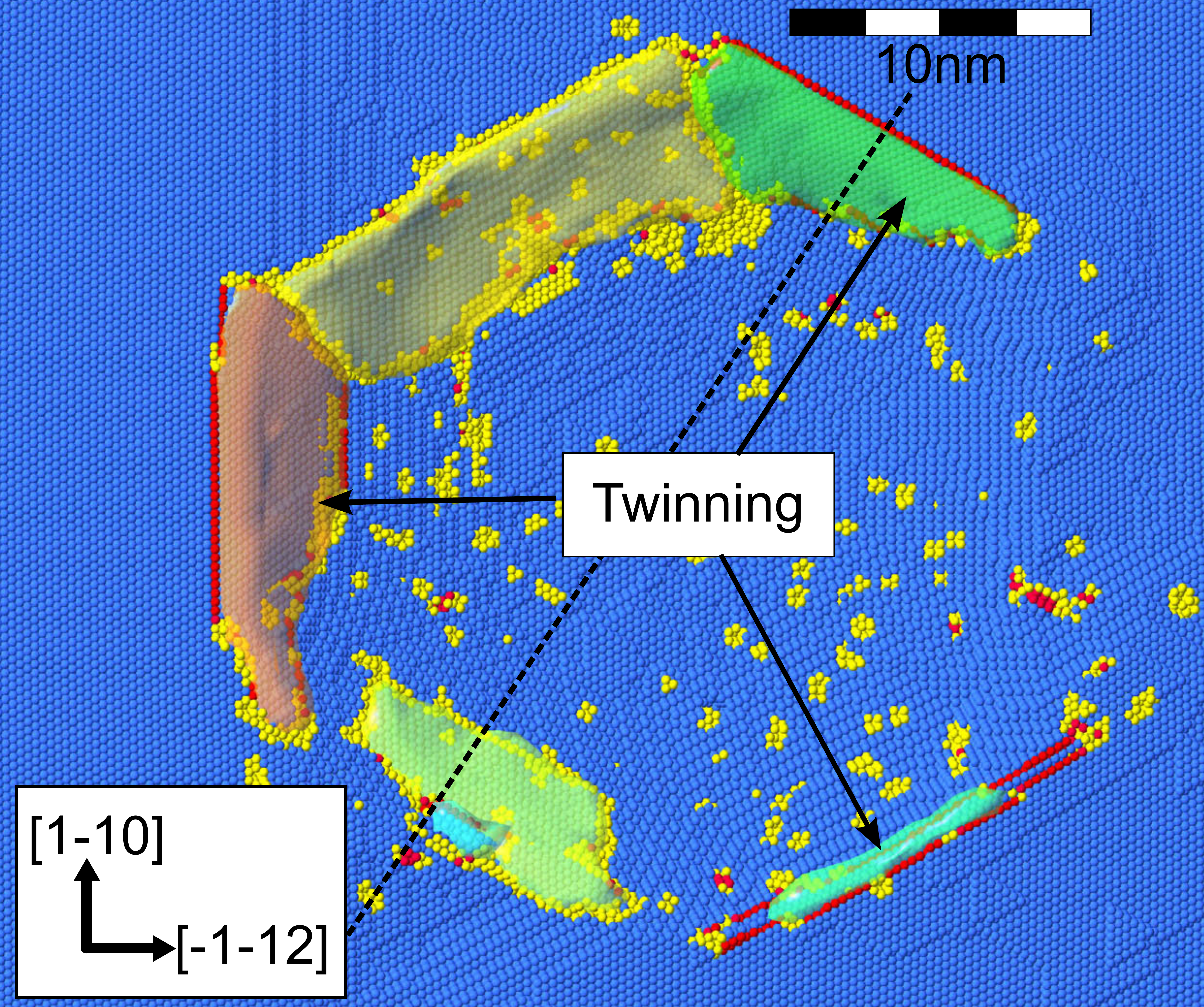}}
		\end{minipage}
	\caption{Twinning and grain rotation in the outer rim of the indenter imprint. \protect\subref{fig:GrainsUnderIndenterA} A cross-sectional cut showing both a grain and a deformation twin. \protect\subref{fig:GrainsUnderIndenterB} Atoms in the vicinity of grain/twin-boundaries (colored in red) and atoms at the free surface are excluded from further analysis steps. \protect\subref{fig:GrainsUnderIndenterC} Deformation twins and subgrains are aligned three-fold symmetrically on the (111) plane, although one grain is missing. The dashed black line indicates the cutting plane of \protect\subref{fig:GrainsUnderIndenterA} and \protect\subref{fig:GrainsUnderIndenterB}. Figures are created from the simulation with $r=\SI{16}{\nano\meter}$ and $d=0.5r$.}
	\label{fig:GrainsUnderIndenter}
\end{figure}
A size effect has been observed in the simulations that had to be accounted for in order to properly characterize the dislocation microstructure and the associated energies.
For indenter sizes of $r=\SI{16}{\nano\meter}$ and larger, the formation of deformation twins and grain rotation is observed in the outer rim of the plastic zone, presumably due to the high dislocation densities in this region.
An example of such a configuration is given in Fig.~\ref{fig:GrainsUnderIndenterA}.
The crystal lattices in these grains are differently oriented than the bulk and they are either bounded by several layers of atoms classified as defects or by single layers of atoms in HCP configuration.
The first case indicates the presence of a grain-boundary, the latter one a coherent twin-boundary.
Alternating grain-boundaries and deformation twins are found to be arranged in a 3-fold symmetrical pattern around the indenter imprint as shown in Fig.~\ref{fig:GrainsUnderIndenterC}, although the third segment of a rotated grain is missing.
Similar patterns are observed in other samples as well.
The mechanisms causing these grain rotations will not to be discussed here further; however, it is necessary to consider them in the analysis of the dislocation density and the associated energy.
Since grain boundaries, as well as the free surfaces, posses a higher energy compared to the bulk crystal, these regions are excluded from the following analysis and only the interior of the crystal (see Fig.~\ref{fig:GrainsUnderIndenterB}) is considered as the volume for the assessment of dislocation densities and free energies.

\floatsetup[figure]{style=plain,subcapbesideposition=top}
\begin{figure}[tbp]
	\centering
		\sidesubfloat[]{\label{fig:fullSamplesDensitiesA}\includegraphics[width=0.45\textwidth]{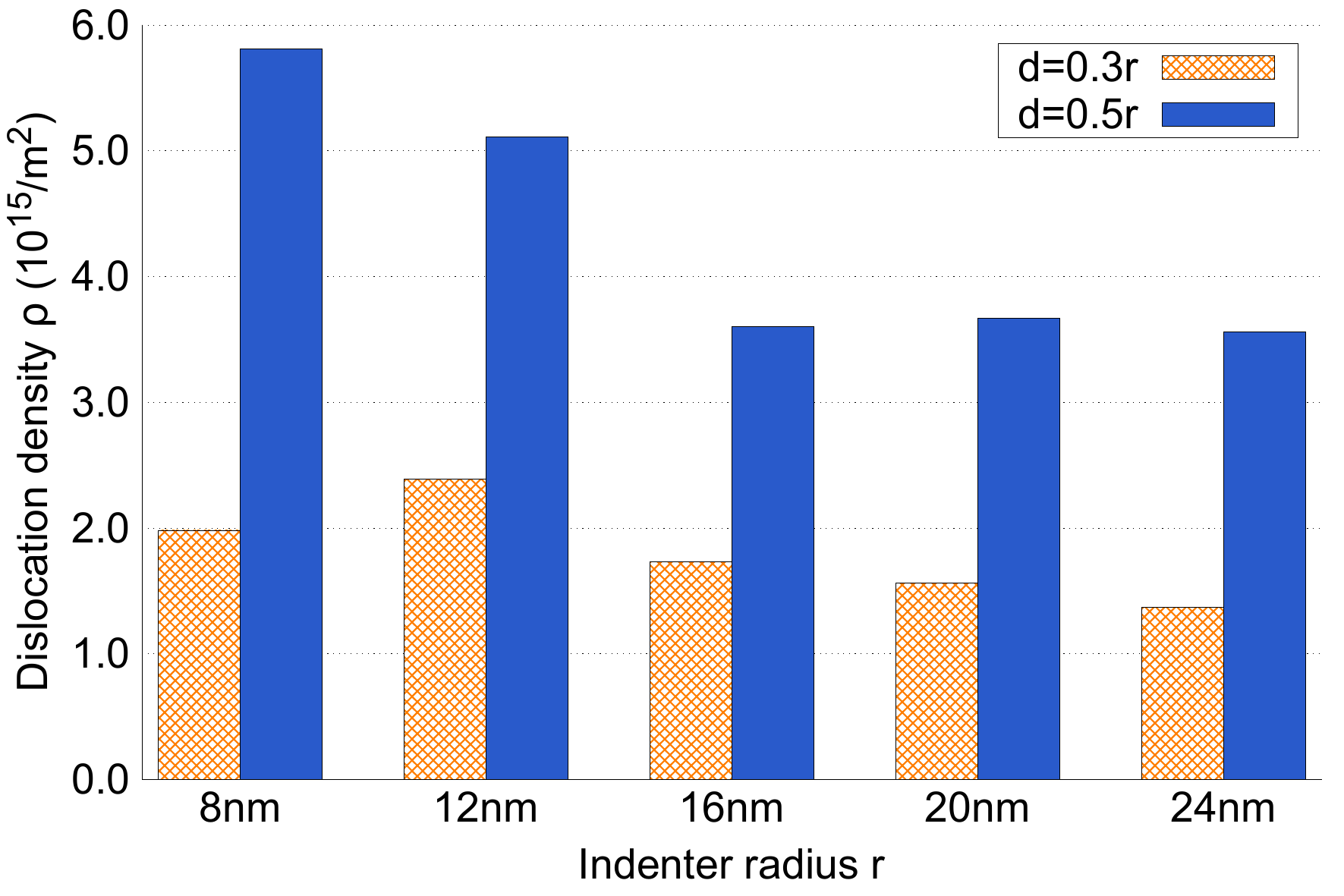}}
		\sidesubfloat[]{\label{fig:fullSamplesDensitiesB}\includegraphics[width=0.45\textwidth]{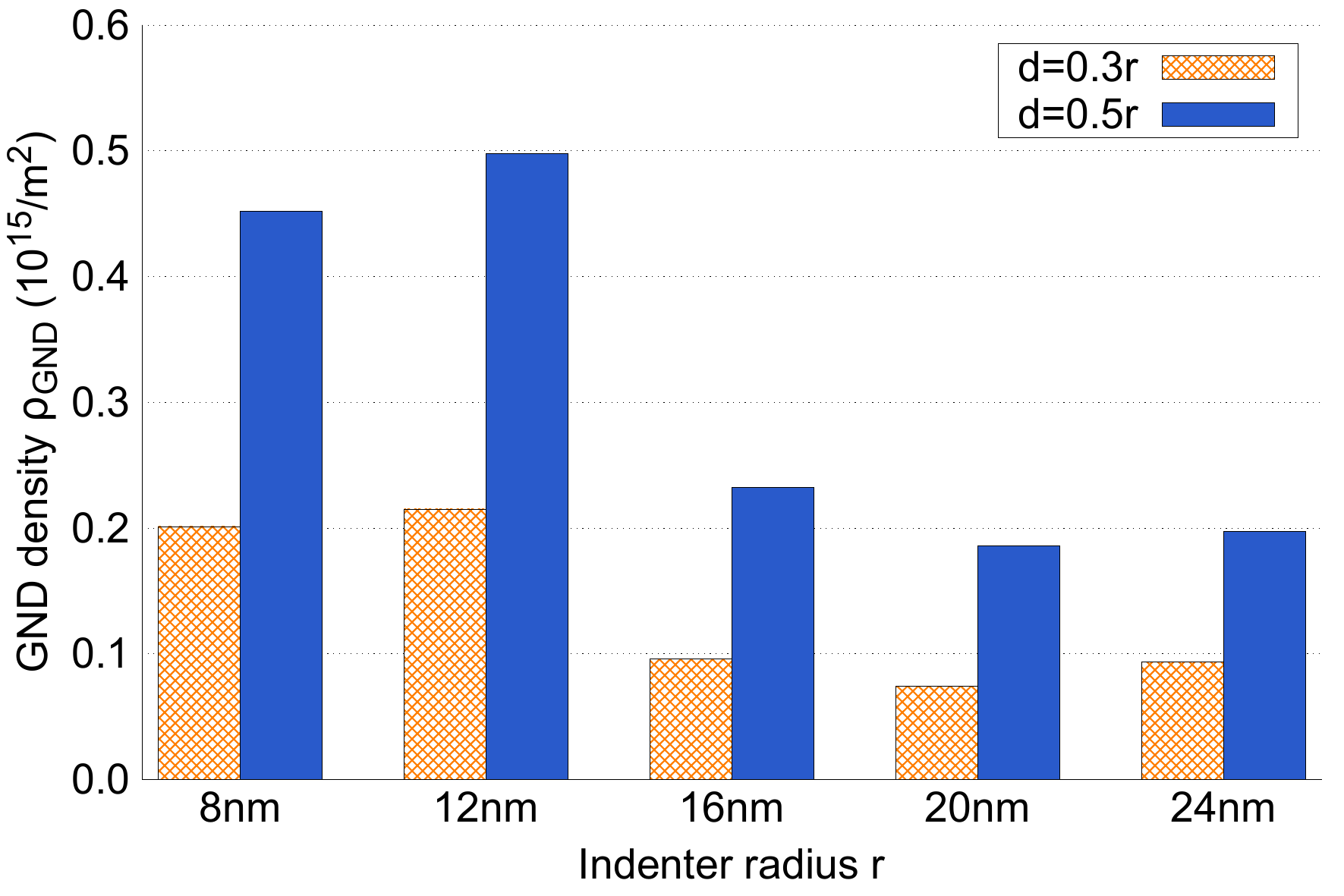}}\\
		\sidesubfloat[]{\label{fig:fullSamplesDensitiesC}\includegraphics[width=0.45\textwidth]{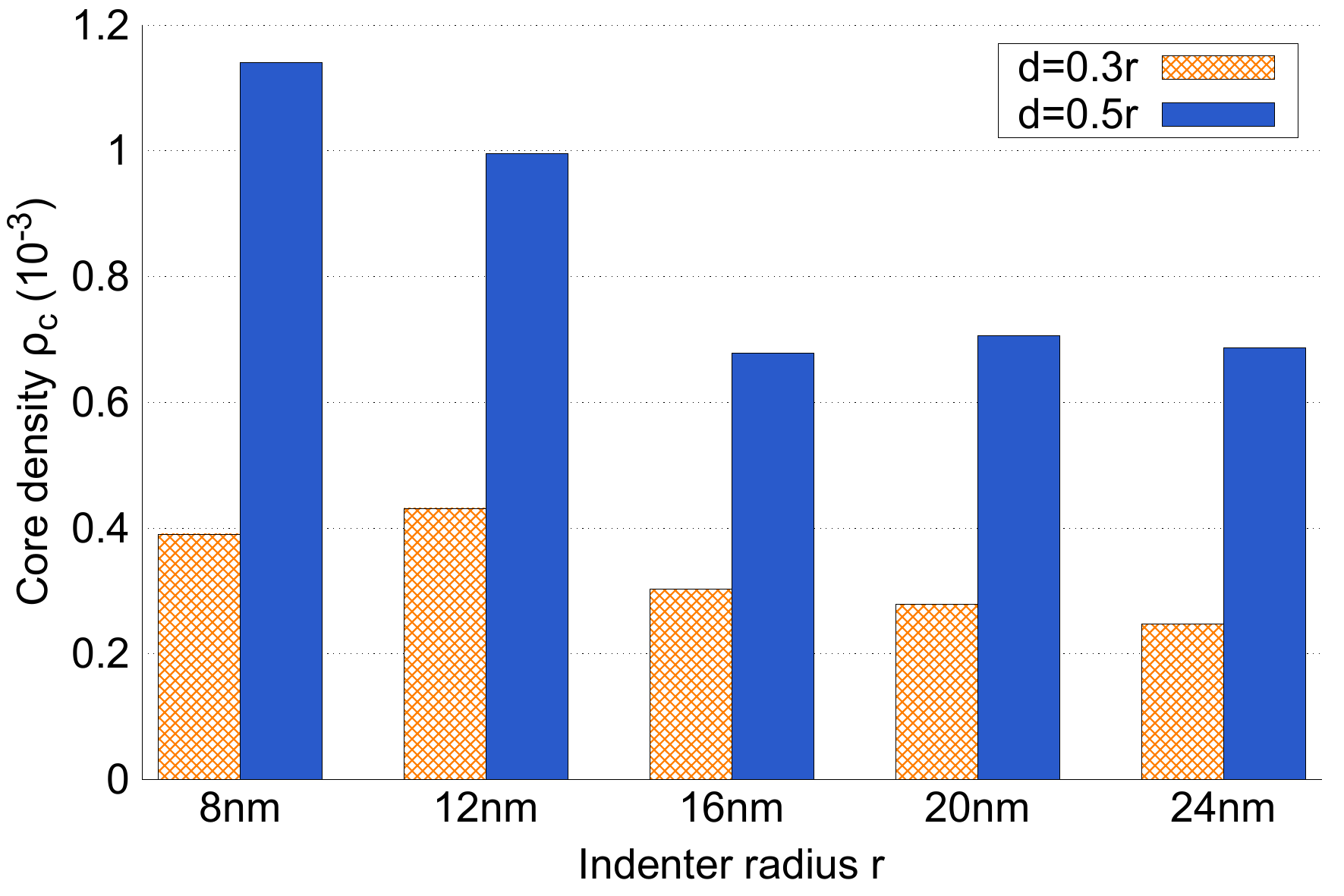}}
		\sidesubfloat[]{\label{fig:fullSamplesDensitiesD}\includegraphics[width=0.45\textwidth]{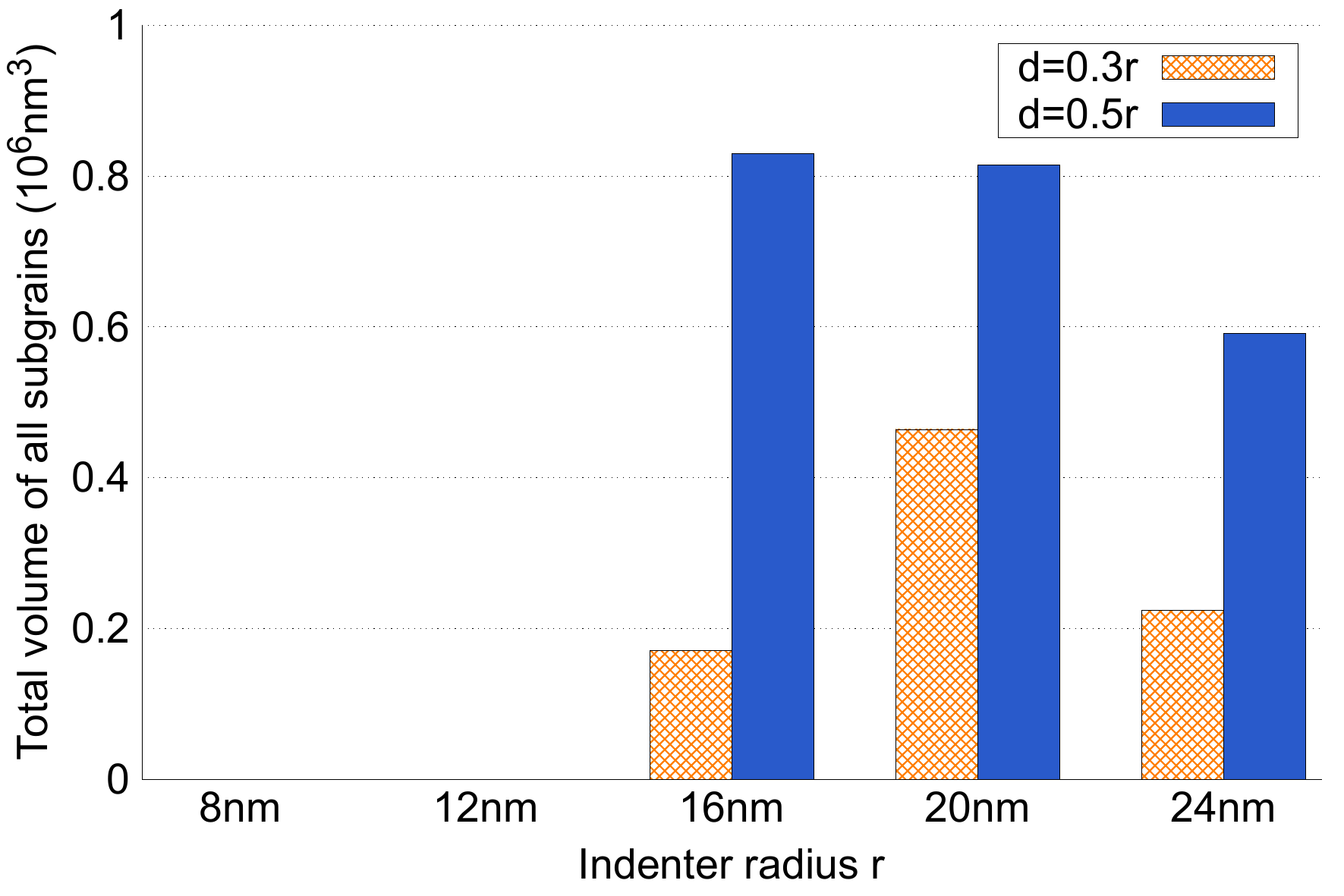}}\\
	\caption{Dislocation densities and the volume of subgrains in the samples.}
	\label{fig:fullSamplesDensities}
\end{figure}

Using the equations given in section~\ref{sec:disDens}, the dislocation densities $\rho$, $\rho_{\mathrm{GND}}$ and $\rho_c$ are directly obtained from the identified dislocation networks within these volumes.
The results are shown in Fig.~\ref{fig:fullSamplesDensities}, including the measured subgrain volumes.
Comparing the dislocation densities of different sample sizes, both $\rho$ and $\rho_{\mathrm{GND}}$ are significantly higher for the case of the two smallest samples, which can be related to the lack of grain rotation and deformation twinning in these samples.
In the larger samples, deformation is partially accommodated by grain rotation and not by dislocation movement, resulting in lower dislocation densities in the volume.
Considering only the three largest indenter radii, approximately constant densities are measured for equal indentation depths.
However, the GND density $\rho_{\mathrm{GND}}$ is at least one order of magnitude smaller than the dislocation density $\rho$ and thus statistically stored dislocations are dominating, whereas the ratio between the values is nearly constant.
This is in agreement with an earlier study of nanoindentation with a spherical tip in copper \cite{Begau2012} where a linearly increase in both the total dislocation $\rho$ and the GND density $\rho_{\mathrm{GND}}$ with the indentation depth in the total simulation volume has been observed.

\begin{figure}[tbp]
	\centering
		 \includegraphics[width=0.50\textwidth]{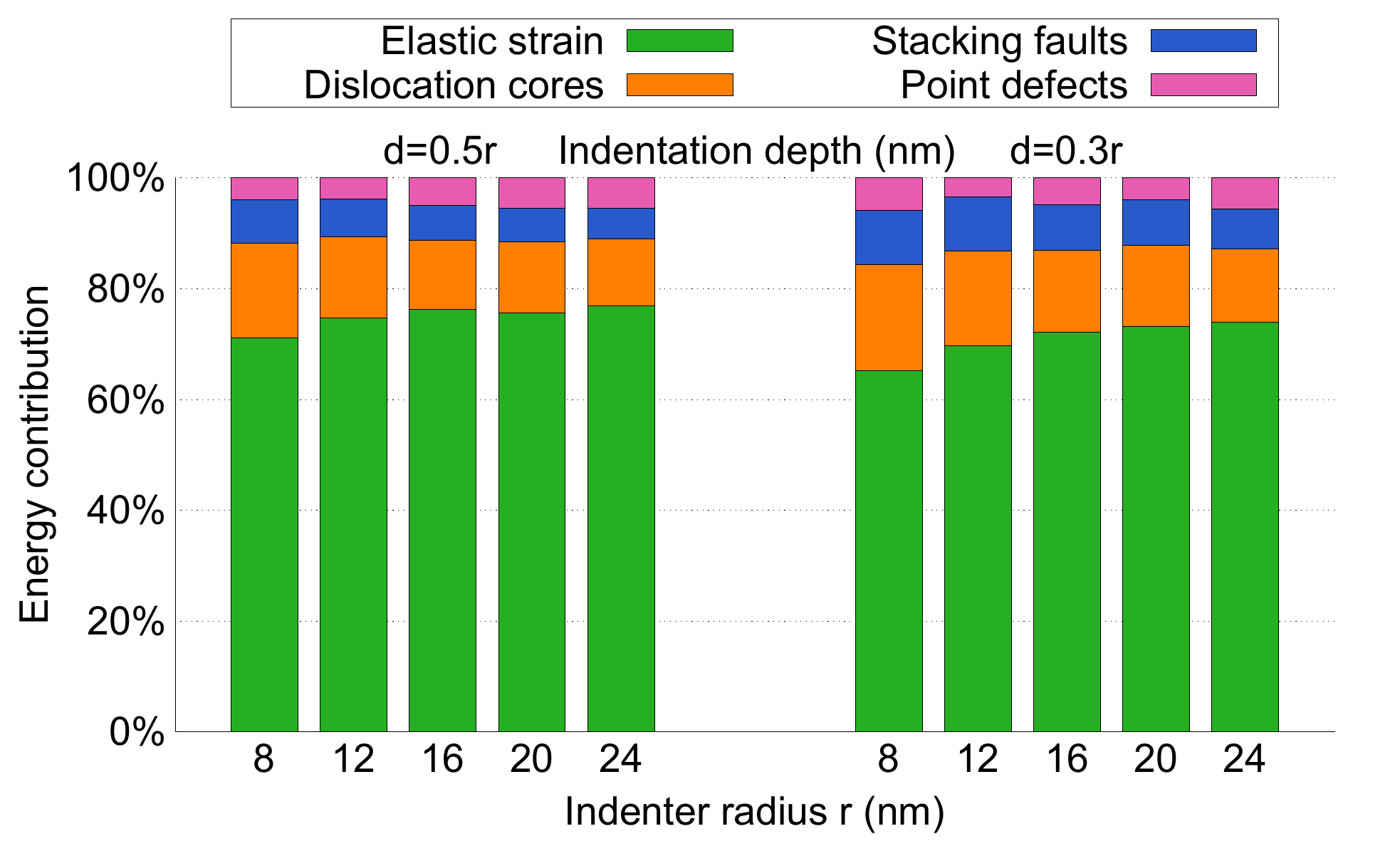}
	\caption{Distribution of energy contributions of different defect entities and of elastic strain.}
	\label{fig:DistributionEnergy}
\end{figure}

\begin{figure}[tbp]
	\centering
		\includegraphics[width=0.5\textwidth]{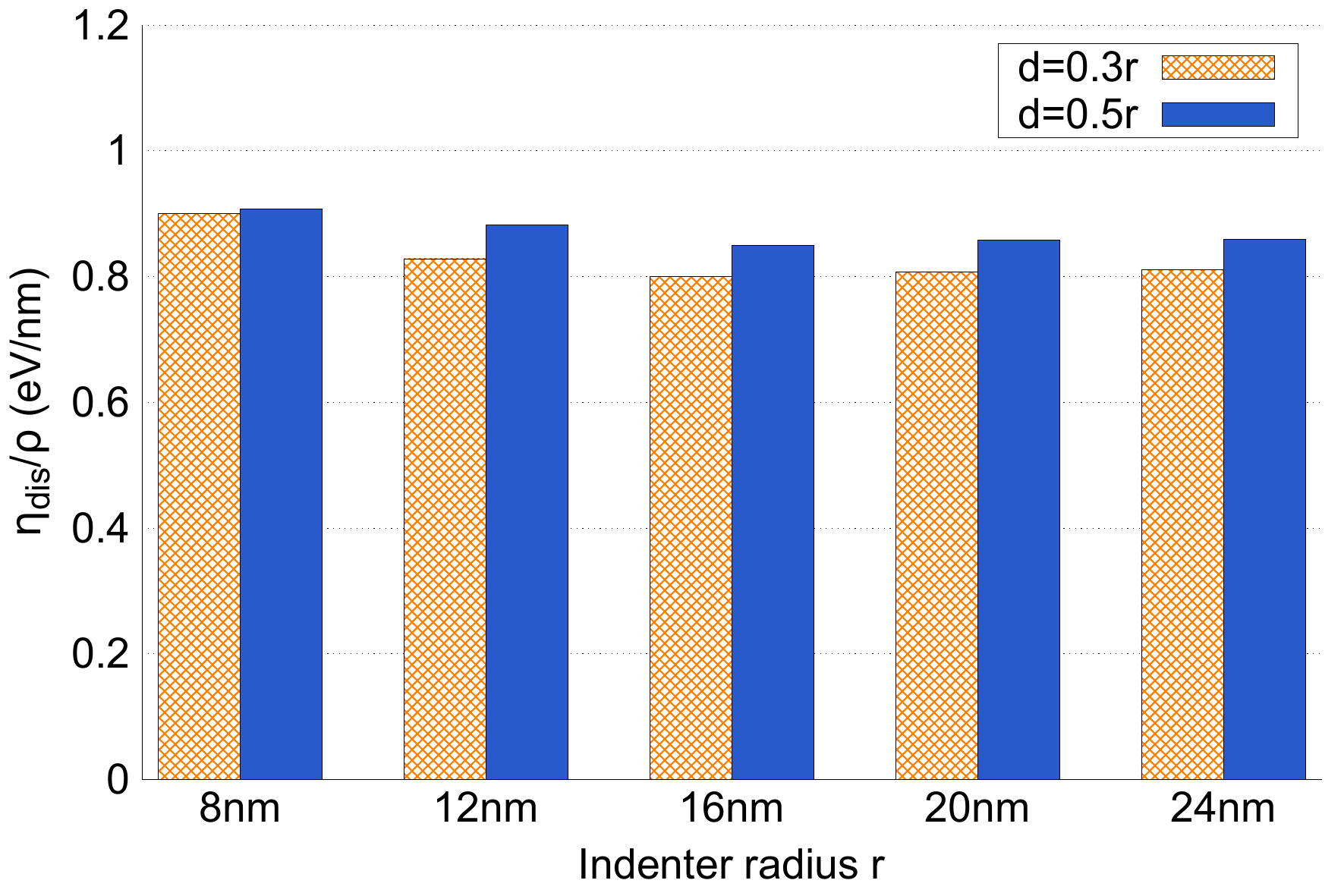}
	\caption{Average energy per unit length of dislocation.}
	\label{fig:energyDisCores}
\end{figure}

The precise measurements of the dislocation densities provide the possibility to relate them to the free energy.
As introduced before, each atom is assigned to a class -- crystal, dislocation core, stacking fault or another defect entity -- and thus it is possible to measure the free energy of each of these classes individually as the sum of $\Delta E$ of all associated atoms.
In Fig.~\ref{fig:DistributionEnergy}, the relative distribution of energies among these defect types is given, showing that the highest partition of energy, on the order of 70\%, is stored in the perfect crystal in the form of elastic energy for all samples.
Dislocation cores and stacking faults contain most of the remaining energy and only a small portion is attributed to point defects.
This result proves that the free energy in the simulations is almost exclusively linked to dislocations, either stored in the dislocation cores itself or in elastic strain caused by the far reaching elastic fields of dislocations.
In the following, the relation of these energies to the dislocation densities is discussed in detail.
First of all the, the core energy density is related to the dislocation density $\rho$, for which a linear relation is to be expected.
The energy density of dislocation cores $\eta_{\mathrm{dis}}$ is defined here as
\be
\eta_{\mathrm{dis}}=(E_{\mathrm{core}}+E_{\mathrm{SF}})/V
\ee
with $E_{\mathrm{core}}$ and $E_{\mathrm{SF}}$ denoting the total energy of dislocation cores and stacking faults.
Since stacking faults are immediately present in dislocations in FCC and store a part of their energy, these two energy contributions are combined in order to approximate the total energy of dislocation cores.
The average energy per unit length of dislocation shown in Fig.~\ref{fig:energyDisCores}, computed as $\eta_{\mathrm{dis}}/\rho$, is basically constant for all simulations with a value of about \SI{0.8}{\electronvolt\per\nano\meter} .
From this results it can be concluded that the total energy of dislocation cores is indeed scaling proportionally to the dislocation density $\rho$ and that, therefore, the core energy cannot be related on the GND density, since the ratio $\rho/\rho_{\mathrm{GND}}$ is not constant for all simulations (cf.~Fig.~\ref{fig:fullSamplesDensities}).

\begin{figure}[tbp]
	\centering
		\sidesubfloat[]{\label{fig:elasticEnergyPerDisA}\includegraphics[width=0.45\textwidth]{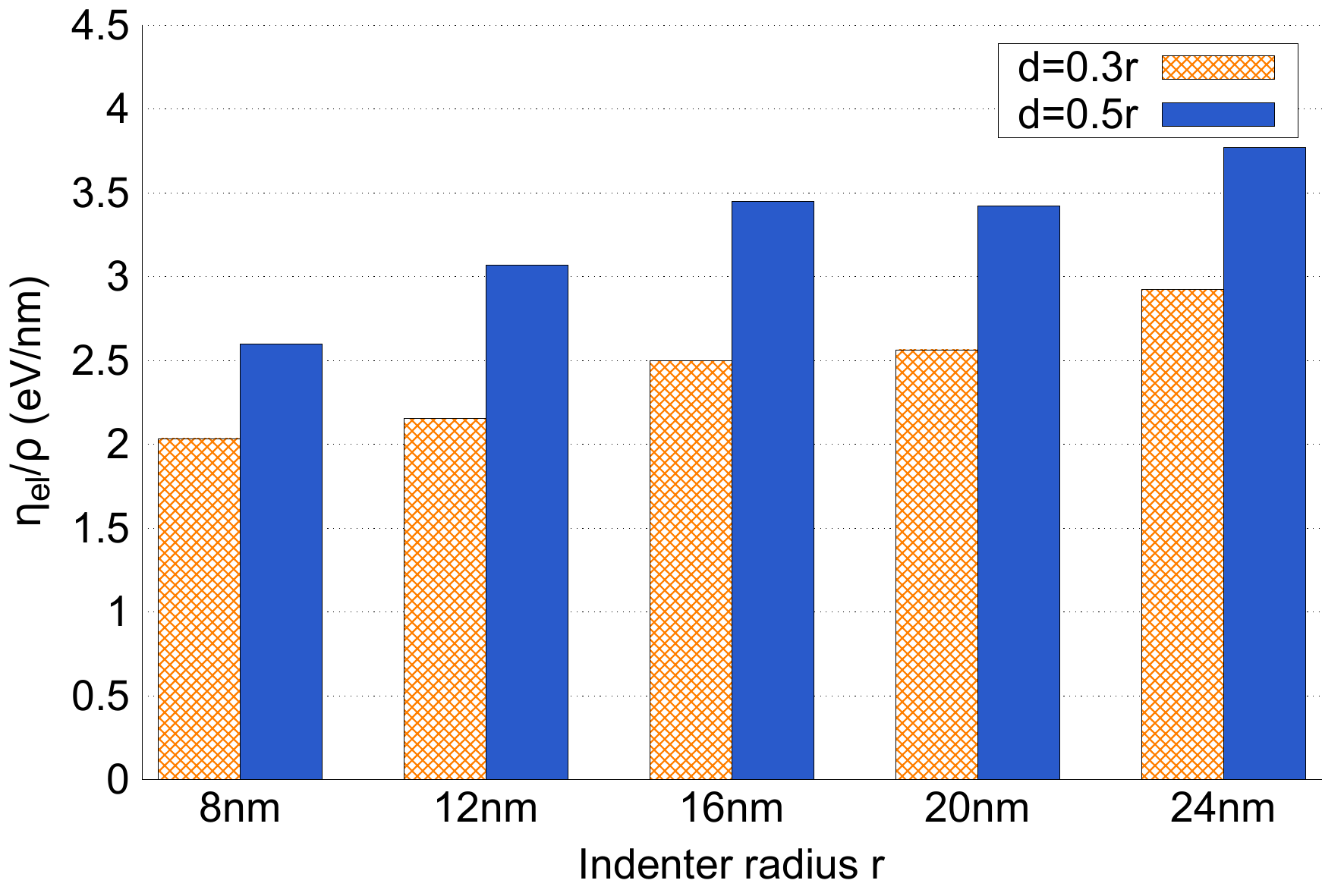}}
		\sidesubfloat[]{\label{fig:elasticEnergyPerDisB}\includegraphics[width=0.45\textwidth]{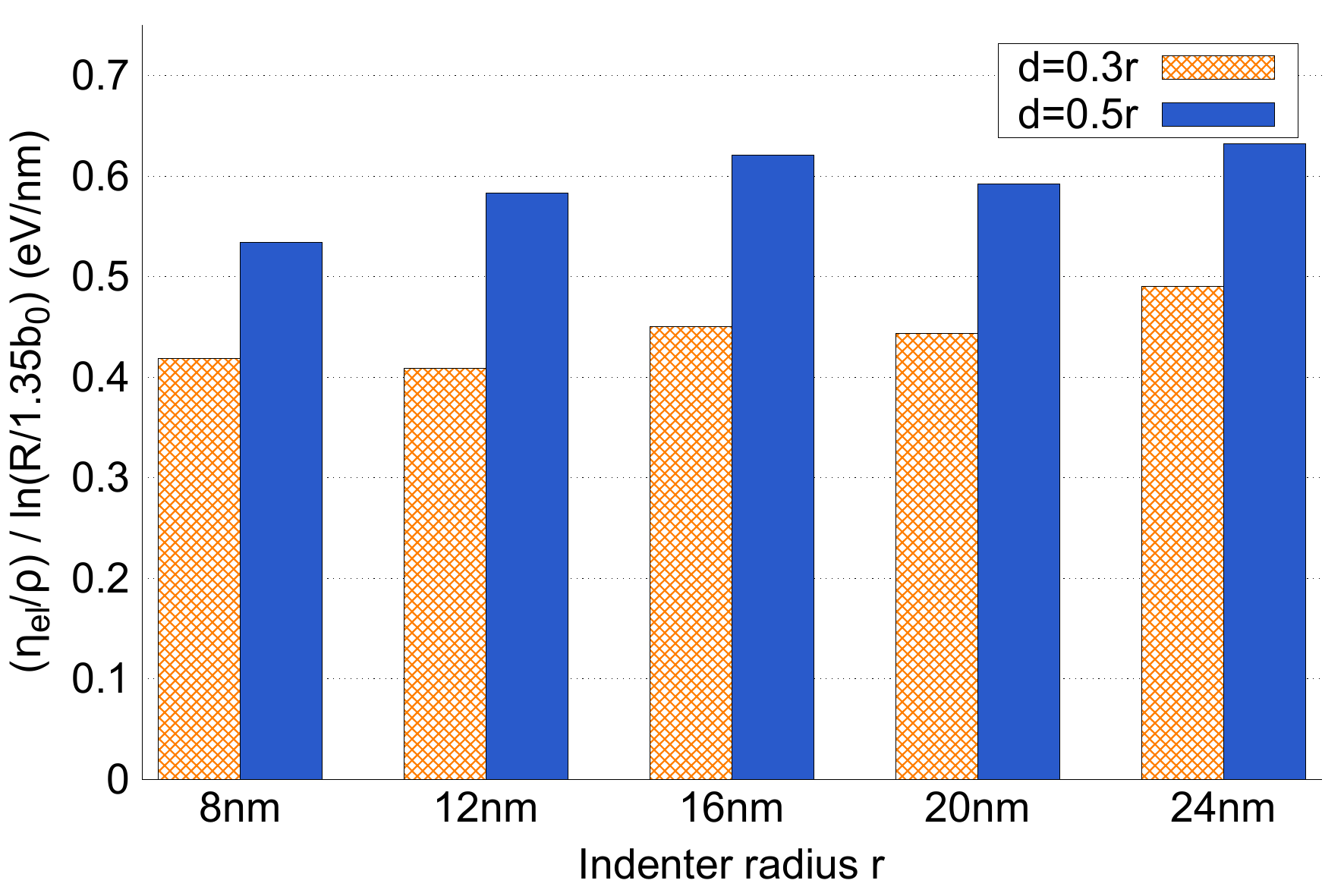}}\\
		\sidesubfloat[]{\label{fig:elasticEnergyPerDisC}\includegraphics[width=0.45\textwidth]{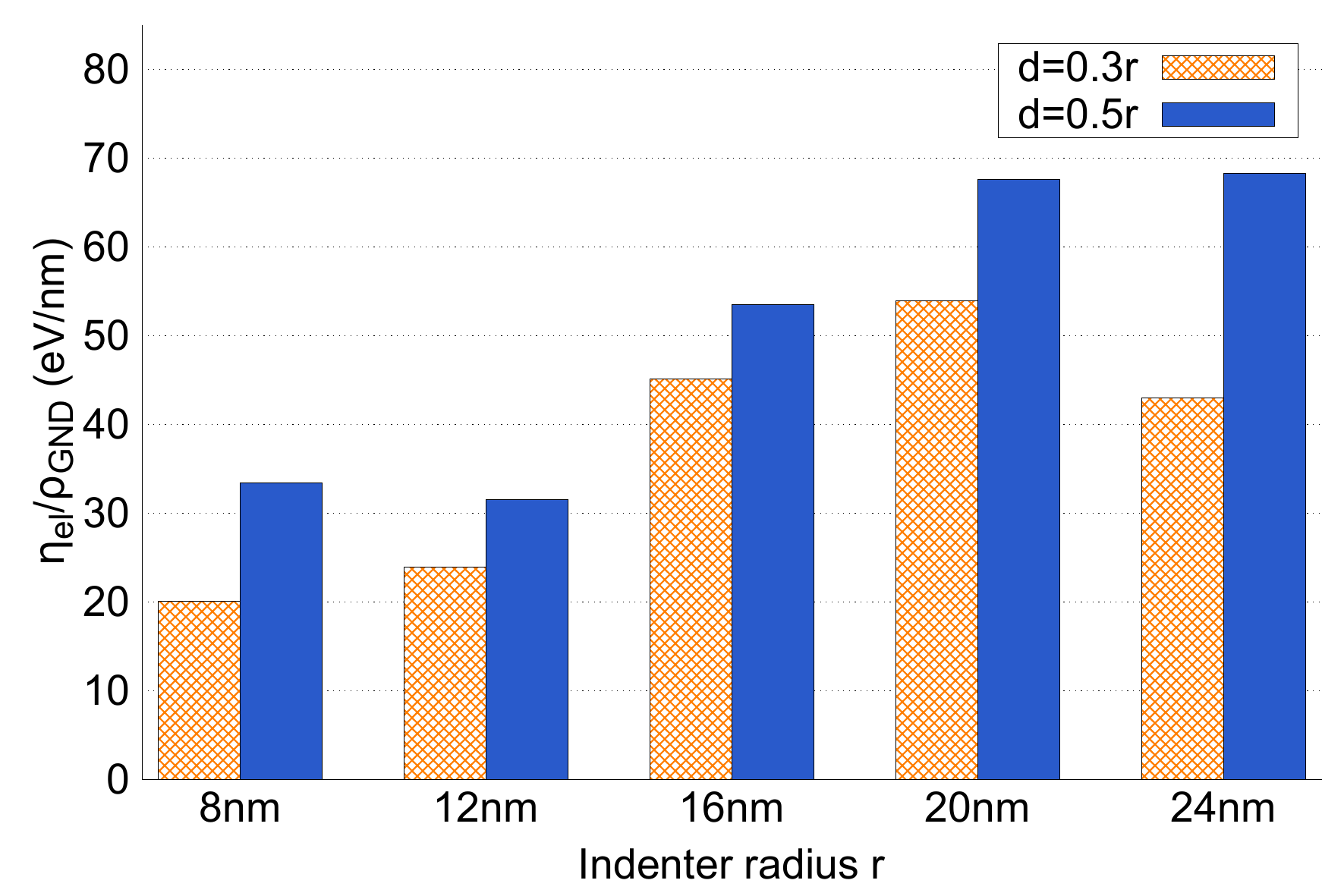}}
		\sidesubfloat[]{\label{fig:elasticEnergyPerDisD}\includegraphics[width=0.45\textwidth]{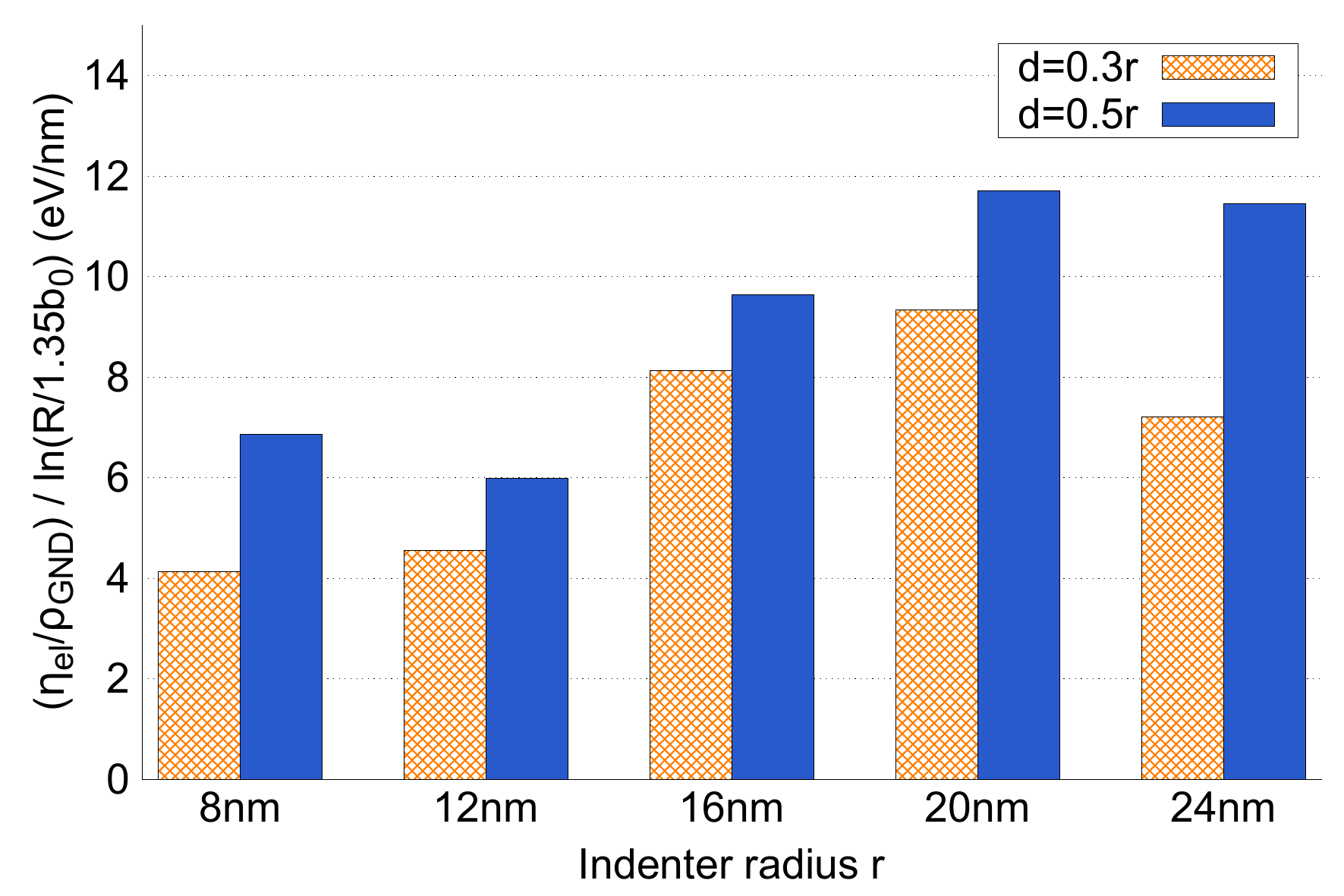}}\\
	\caption{Relation of elastic energy density to $\rho$ \protect\subref{fig:elasticEnergyPerDisA} and $\rho_{\mathrm{GND}}$ \protect\subref{fig:elasticEnergyPerDisC}. A size dependent correction factor for both these values is included in \protect\subref{fig:elasticEnergyPerDisB} and \protect\subref{fig:elasticEnergyPerDisD}.}
	\label{fig:elasticEnergyPerDis}
\end{figure}

However, the energy of dislocation cores is only a small fraction of the total energy stored during plastic deformation as shown in Fig.~\ref{fig:DistributionEnergy}, whereas the majority is stored in elastic strains.
Hence, elastic energy needs to be related to dislocation densities to establish a free energy function of the dislocation microstructure.
The relation of elastic energy density $\eta_{\mathrm{el}} = E_{\mathrm{el}}/V$ to dislocation density $\rho$ is plotted in Fig.~\ref{fig:elasticEnergyPerDisA} and to GND density $\rho_{\mathrm{GND}}$ in Fig.~\ref{fig:elasticEnergyPerDisC}.
A noticeable growth in the energy per unit length of dislocation is observed with the simulation box size $s$, which is in stark contrast to the dislocation densities $\rho$ and $\rho_{\mathrm{GND}}$ that converge to approximately constant values for larger simulation volumes (Fig.~\ref{fig:fullSamplesDensities}).
In case of GNDs, the trend of an increasing energy per unit length is visible as well, but it is significantly stronger scattered.
Based on this observation, a non-linear dependence between the free energy, the dislocation densities and the sample volume must be assumed.
Such a relation is in fact to be expected according to the classical model of elastic energy of a single dislocation as given in Eq.~\ref{Eq:ElasticEnergy}.
This model includes a logarithmic factor based on a characteristic length and the dislocation core radius, which accounts for the long-ranged stress fields of dislocations.
In Figs.~\ref{fig:elasticEnergyPerDisB} and~\ref{fig:elasticEnergyPerDisD} such a factor has been included in the relation of the measured elastic energy and the dislocation densities, using the lateral size of the simulation box $s$ as the parameter $R$ and setting the size of the dislocation core radius as $r_0=c\cdot b_0=\SI{0.382}{\nano\meter}$.
The factor $c=1.35$, used to relate the magnitude of the Burgers vector to the dislocation core radius, has been derived from the quantities $V$, $\lambda$ and $\rho_c$ which are measurable within the framework of the simulations according to Eqs.~\ref{Eq:lb2} and \ref{Eq:rho_c_by_atoms}.
The value of $c=1.35$ is the mean of all simulations, with a standard deviation of 0.037.
\floatsetup[figure]{style=plain,subcapbesideposition=top}
After inclusion of this correction factor, the trend between the elastic energy and the dislocation density $\rho$ is becoming evident.
As shown in Fig.~\ref{fig:elasticEnergyPerDisB}, the energy per unit length of dislocation is on a comparable level for equal indentation depths without showing a clear increase in energy with the size anymore, although the results are considerably scattered.
In contrast to these results, no trend is visible comparing the GND density with the energy density as shown in Fig.~\ref{fig:elasticEnergyPerDisD}.
This is a first clear indicator that the free energy of the dislocation density does not only depend on the GND density, but also on the total dislocation density, which seems be to the most relevant quantity.

\begin{figure}[tbp]
	\centering
		\includegraphics[width=0.40\textwidth]{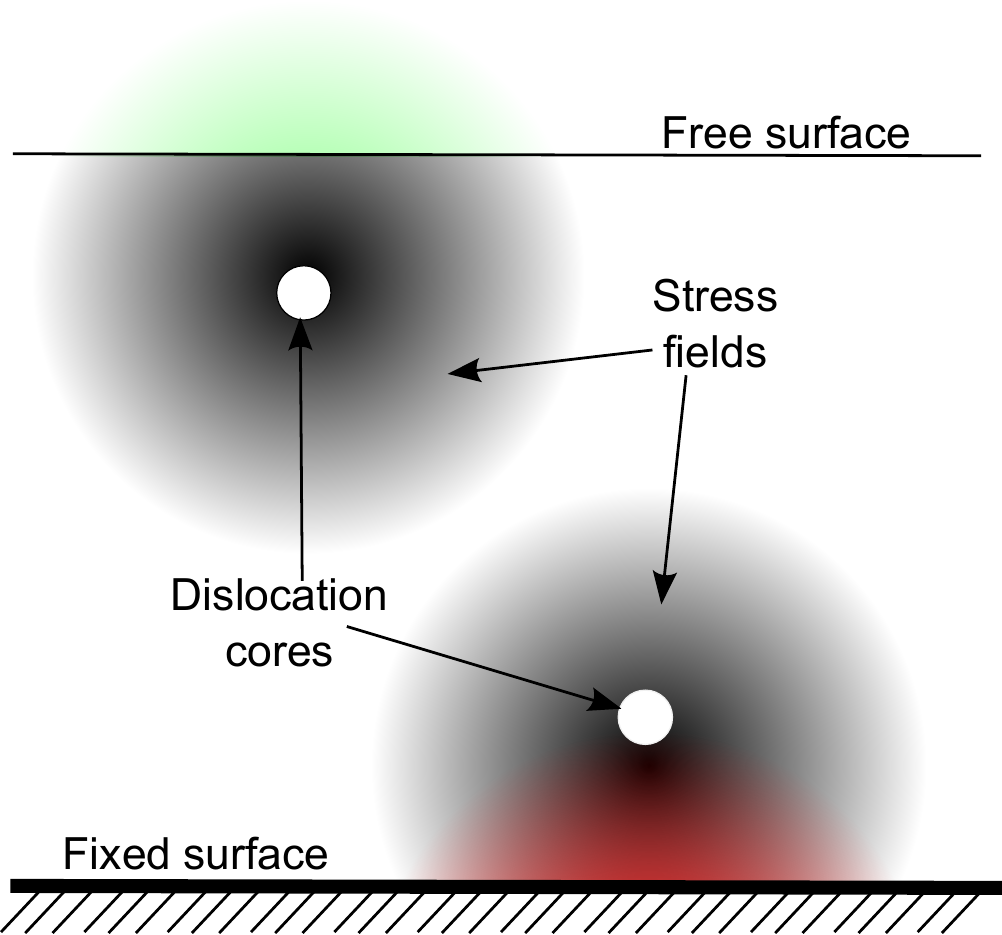}
	\caption{Illustration of a possible surface effect. Stress and strain fields around dislocation cores are directly related to the elastic energy. At the free surface, this stress is partly reduced (green shaded area), whereas it is concentrated at a fixed boundary (red shaded area).}
	\label{fig:StressFields}
\end{figure}

According to Fig.~\ref{fig:elasticEnergyPerDisB}, similar energy densities relative to $\rho$ are measured for equal indentation depths, whereas there is a strong discrepancy between the two different indentation depths of $d=0.3r $ and $d=0.5r$.
These effects may be related to the different local GND densities for the two indentation depths, but other explanations are possible as well, e.g.\ it can be considered as a surface effect, originating in the stress fields of dislocations that can be linked to Eq.~\ref{Eq:ElasticEnergy} as well.
As visible in Fig.~\ref{fig:DisloationNetworks}, more dislocations are located towards the bottom of the simulation box for the cases of $d=0.5r$, where several atomic layers are fixed by the applied boundary conditions to avoid rigid body motion in the simulation box during indentation.
This configuration is influencing the stress and strain fields around dislocation cores as illustrated in Fig.~\ref{fig:StressFields}.
At the free surface the stress and strain fields are reduced, such that less elastic energy is stored in total for dislocations near the surface.
At the fixed boundaries instead dislocations are piling up, causing a higher elastic strain energy density in this domain.
This effect is very likely to contribute to the different elastic strain energy densities observed at the different indentation depths; however, it is outside the scope of this work to verify if this effect is solely responsible or if other mechanisms need to be taken into account for an explanation as well.

In conclusion, several observations on the dislocation and elastic energy densities are made on the full sized samples.
Both dislocation densities $\rho$ and $\rho_\mathrm{{GND}}$ converge to fairly constant values at a constant indentation depth relative to the tip radius once a size effect associated with the formation of subgrains and deformation twinning is accounted for.
In contrast to this, the elastic energy density is increasing in larger simulation domains that can be related to the long ranged stress fields of dislocations, similarly to the model used to estimate the energy of single straight dislocations.
This indicates that the free energy of a self-assembled dislocation microstructure in a deformed single crystal can be estimated as the superposition of many single dislocations and thus should scale linearly with the density $\rho$.

However, due to the observed inhomogeneous dislocation distributions, these averaged values for the full samples provide only an incomplete image of the local densities.
Whereas the dislocation density in the plastic zone seems to be saturated, because other deformation mechanism such as subgrain formation are observed, the average core density $\rho_c$ in the total simulation box is only on the order of $10^{-3}$.
Therefore, the assembled atomistic data is analyzed in more detail for multiple smaller subvolumes, thus providing a better sampling of the inhomogeneously distributed dislocations.

\subsection{Spatially resolved analysis of dislocation densities}
\label{sec:Subvolumes}
Dislocation densities are always computed with respect to a certain reference volume on an internal length scale.
This is particularly important for the simulations here, which possess severely inhomogeneously distributions of dislocations.
Typically, models used in strain gradient crystal plasticity as introduced before do not take a minimum volume into account, although the size of the reference volume $V$ is crucial to compute dislocation densities.
Most importantly, the GND density is highly size dependent, since in a sufficiently small volume every dislocation is geometrically necessary and the equation $\rho = \rho_{\mathrm{GND}}$ must hold true.
In contrast, very large reference volumes typically contain many compensating dislocations and thus $\rho \gg \rho_{\mathrm{GND}}$, as shows in the previous section.
This simple observation raises the question if any fundamental difference between $\rho$ and $\rho_{\mathrm{GND}}$ can exist for an energy function or if it is at least dependent on the size of the reference volume or some internal length scale.

\begin{figure}[tbp]
	\centering
		\includegraphics[width=0.90\textwidth]{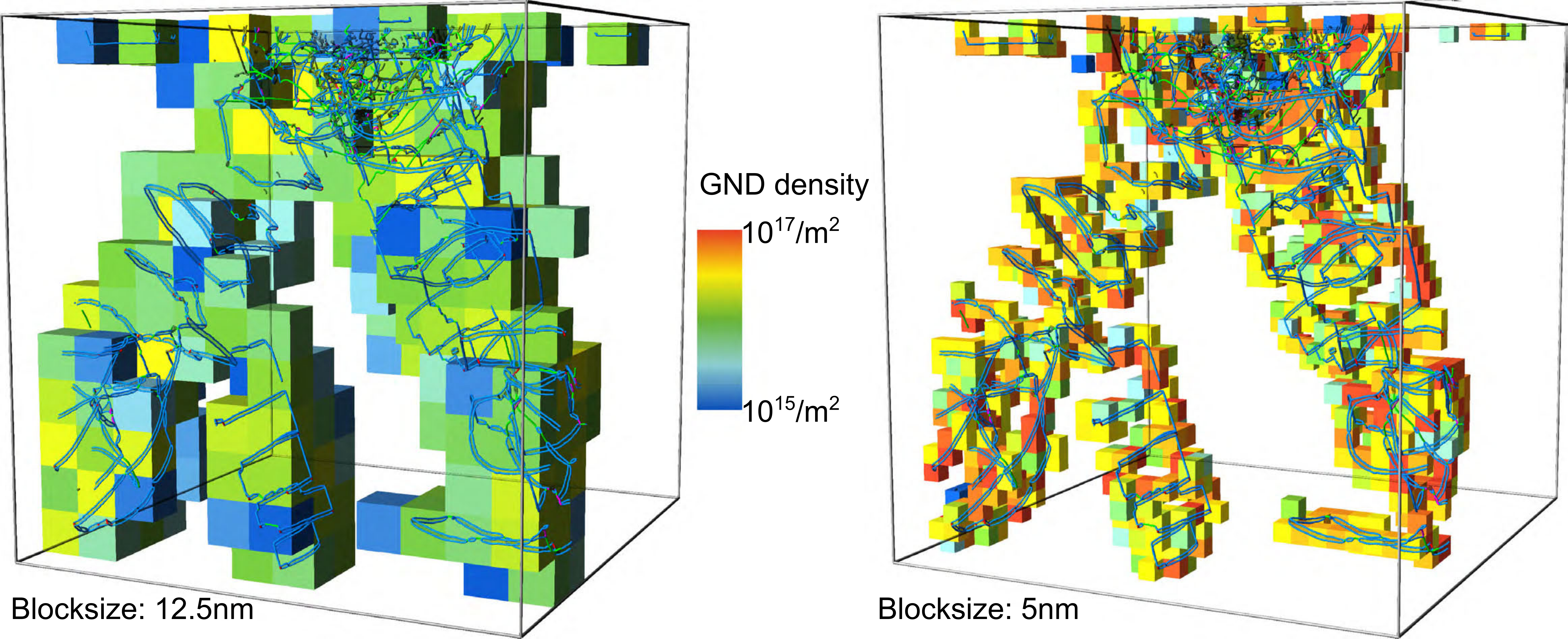}
	\caption{Illustration of the local GND densities for varying reference volumes. The dislocation network is projected on top of the blocks containing a non-zero density.}
	\label{fig:Blocksize}
\end{figure}

Since the simulated volumes are rather large for atomistic studies (up to $150^3$\,\si{\cubic\nano\meter}), the total volume can be subdivided into smaller blocks, still consisting of a sufficiently large number of atoms to measure dislocation densities and energies.
This way, a much larger number of samples to compare free energies and dislocation densities is gathered.
Taking into account especially the dependency of the GND density on the reference volume, four different block sizes are computed for all ten indentation simulations in which cubic blocks of lateral lengths of \SI{5}{\nano\meter}, \SI{6.25}{\nano\meter}, \SI{12.5}{\nano\meter} and \SI{25}{\nano\meter} are used.
An illustration of the construction of these blocks is given in Fig.~\ref{fig:Blocksize}.
This illustration shows another effect of decreasing block sizes, where higher densities are observed in smaller volumes.
Thus, by using different block sizes, a wide range of dislocation densities is measured.

An issue arises for very small blocks, since dislocation cores are of finite volume and may often be located at the interface of multiple blocks.
Therefore, slightly different definitions of dislocation densities are used in this section that are modified in such a way that the contribution of a dislocation is weighted by the number of atoms associated with it, e.g.\ the dislocation density inside a block is defined here as
\be
\rho^{(\mathrm{block})} = \frac{1}{V}\int_{\perp \in V} d\vec{l}_{\perp} \cdot \frac{\mathrm{a_\perp^V}}{\mathrm{a_\perp}} \quad,
\ee
where the total number of atoms in a dislocation core is denoted as $\mathrm{a_\perp}$ and the subset that is inside the volume $V$ as $\mathrm{a_\perp^V}$.
Thus different fractions of a dislocation core can be attributed to multiple blocks by the weighting factor $\mathrm{a_\perp^V}/\mathrm{a_\perp}$.
The definition of the dislocation density tensor $\boldsymbol{\alpha}$ is modified accordingly.

Furthermore, an unphysical effect has been observed using the method of Arsenlis~\&~Parks~\cite{Arsenlis1999} to compute $\rho^{(\mathrm{block})}_{\mathrm{GND}} = \norm{\boldsymbol{\alpha}}$ in very small volumes.
The implemented minimization scheme considers only the existence of Burgers vectors of type $1/2\langle110\rangle$ and $1/6\langle112\rangle$, whereas the dislocation networks in FCC materials contain a small number of other dislocations types, including stair rod and Frank partials, as well.
To create an equal net Burgers vector for these types, a superposition of multiple perfect and partial Burgers vectors might be required which results in an overestimation of the GND density and thus seemingly $\rho_{\mathrm{GND}} > \rho$.
To avoid this problem, the GND density in a block is defined as $\rho^{(\mathrm{block})}_\mathrm{{GND}}:=\mathrm{min}(\norm{\boldsymbol{\alpha}}, \rho^{(\mathrm{block})})$.

Only a subset of all blocks is considered for the analysis of dislocation densities and the free energy.
A block is excluded if it fulfills at least one of the following criteria:
\begin{itemize}
 \item The dislocation density is zero
 \item More than 5\% of the atoms in the block are characterized as free surface of as a grain boundary
 \item The energy stored in point defects, grain boundaries and free surface is larger than 15\% of those in dislocations and stacking faults
\end{itemize}

Due to the large volumes that are free of dislocation, the first criterion excludes the majority of all blocks, especially at small block sizes.
The other two criteria are used to filter those blocks in which the measured energy is biased severely by non-dislocation defects.
In total, from the total 202,258 blocks gathered from all simulations, 8847 are taken into account for comparison.
As blocks containing a significant amount of non-dislocation defects are excluded, the free energy of the blocks can be almost exclusively related to dislocations, either in the form of core energy or elastic energy, thus no distinction of different defect types is required in this section.
In order to provide a comparable quantity for the different block sizes, the free energy in a subvolume is measured as $\Delta E^{\mathrm{av}}_{\mathrm{pot}}$ which is the average $\Delta E_{\mathrm{pot}}$ of all atoms inside the volume.

\floatsetup[figure]{style=plain,subcapbesideposition=top}
\begin{figure}[tbp]
	\centering
		\sidesubfloat[]{\label{fig:GND_rho_energy_subvolumesA}\includegraphics[width=0.46\textwidth]{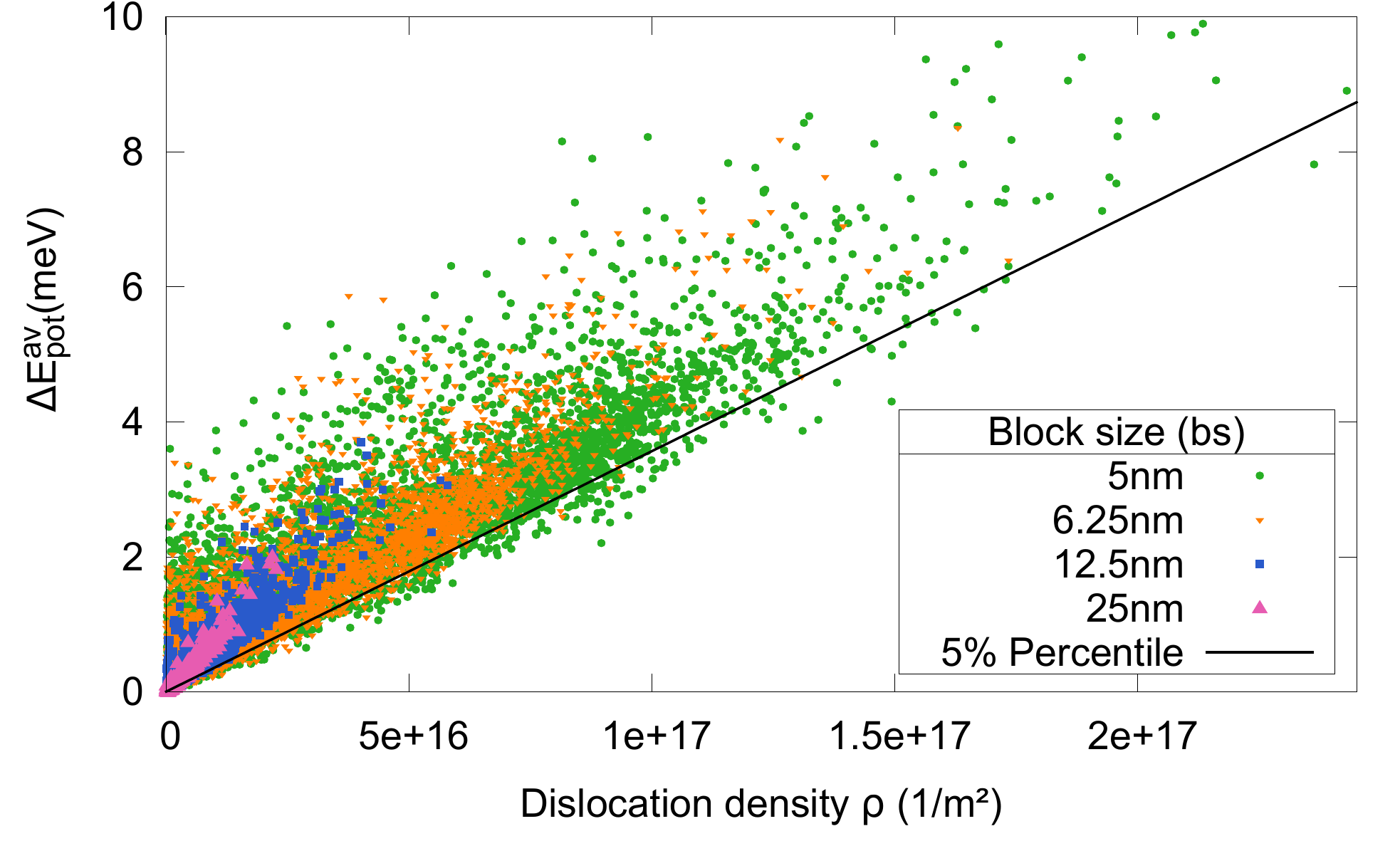}}
		\sidesubfloat[]{\label{fig:GND_rho_energy_subvolumesB}\includegraphics[width=0.46\textwidth]{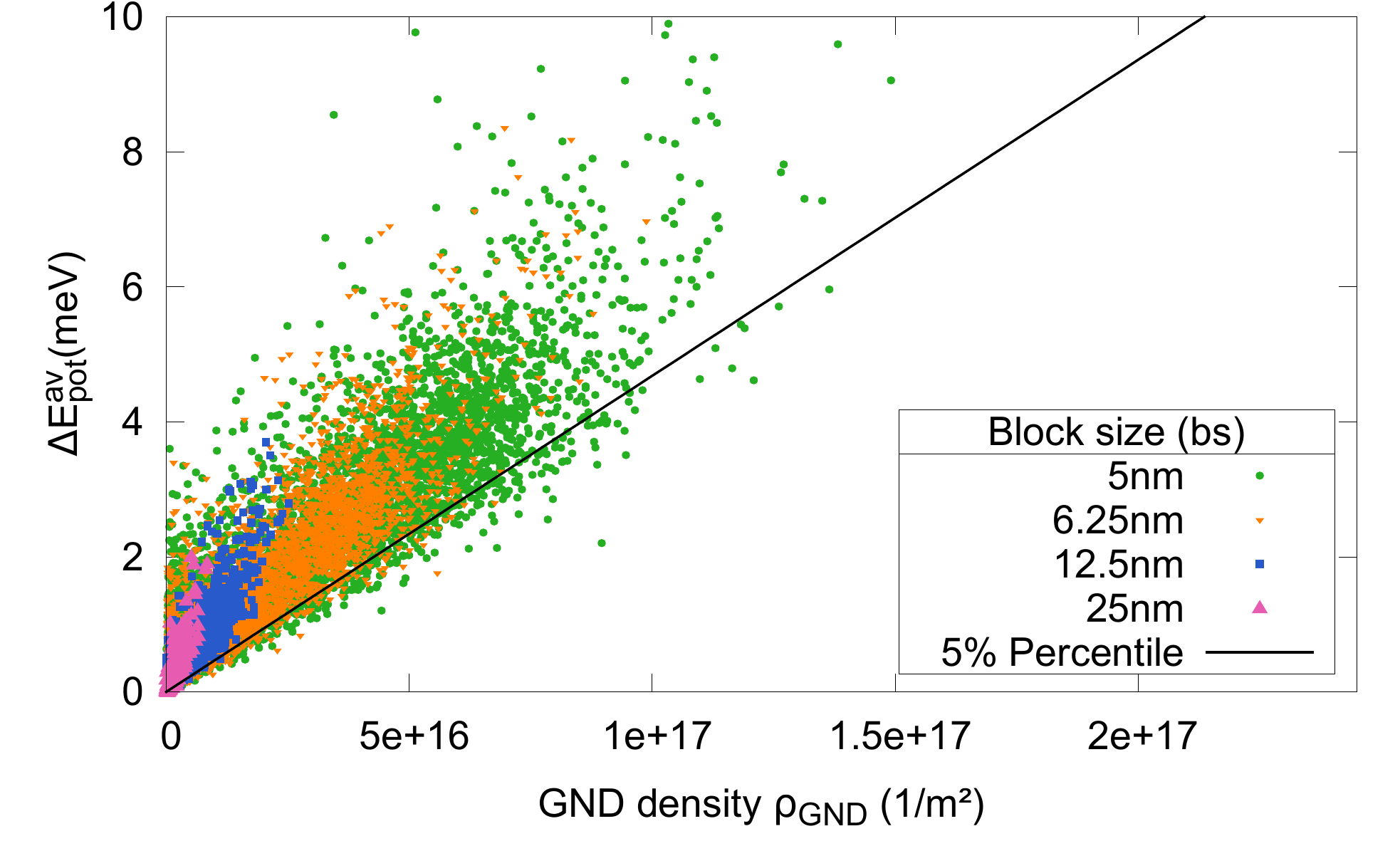}}\\
		\sidesubfloat[]{\label{fig:GND_rho_energy_subvolumesC}\includegraphics[width=0.46\textwidth]{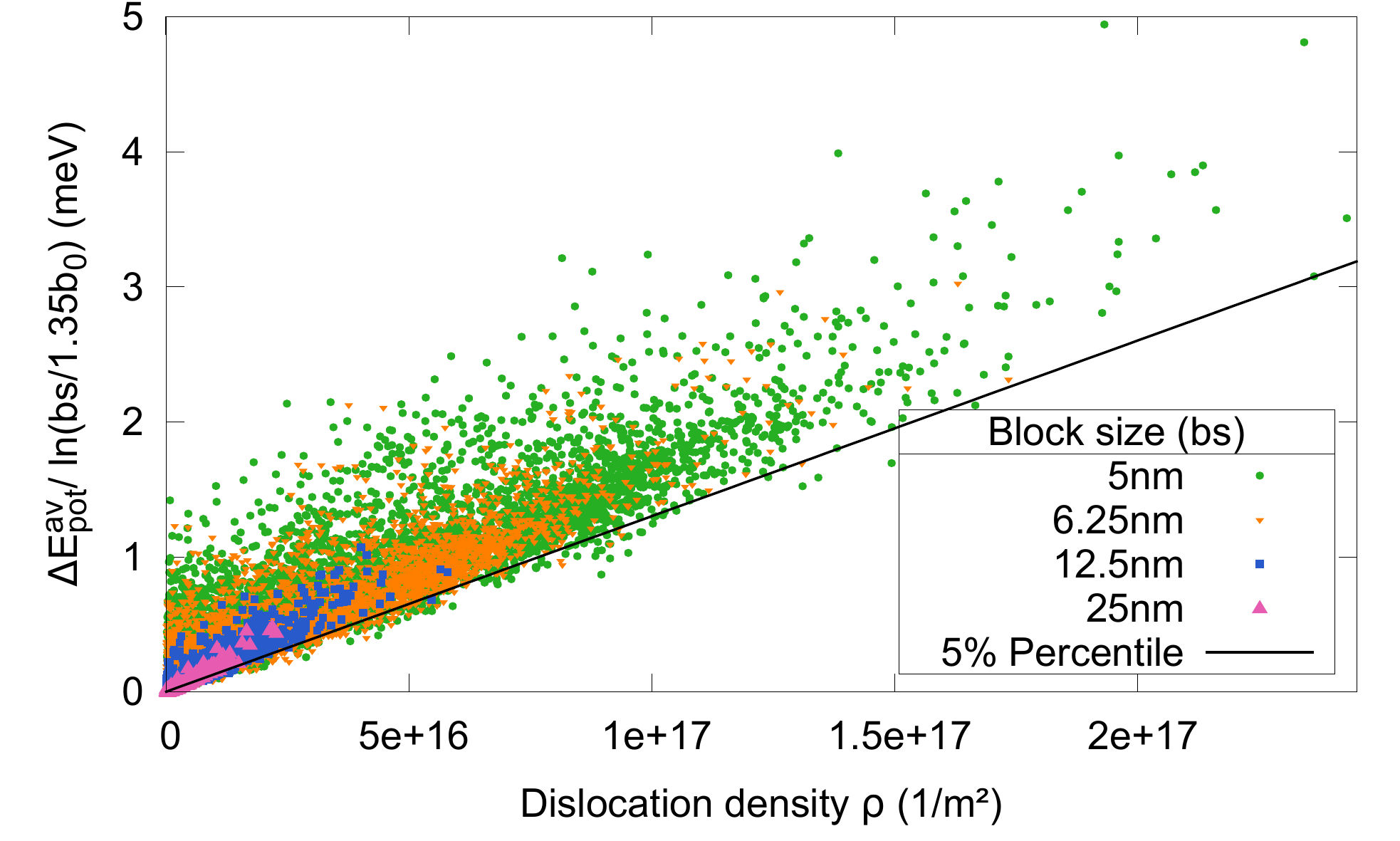}}
		\sidesubfloat[]{\label{fig:GND_rho_energy_subvolumesD}\includegraphics[width=0.46\textwidth]{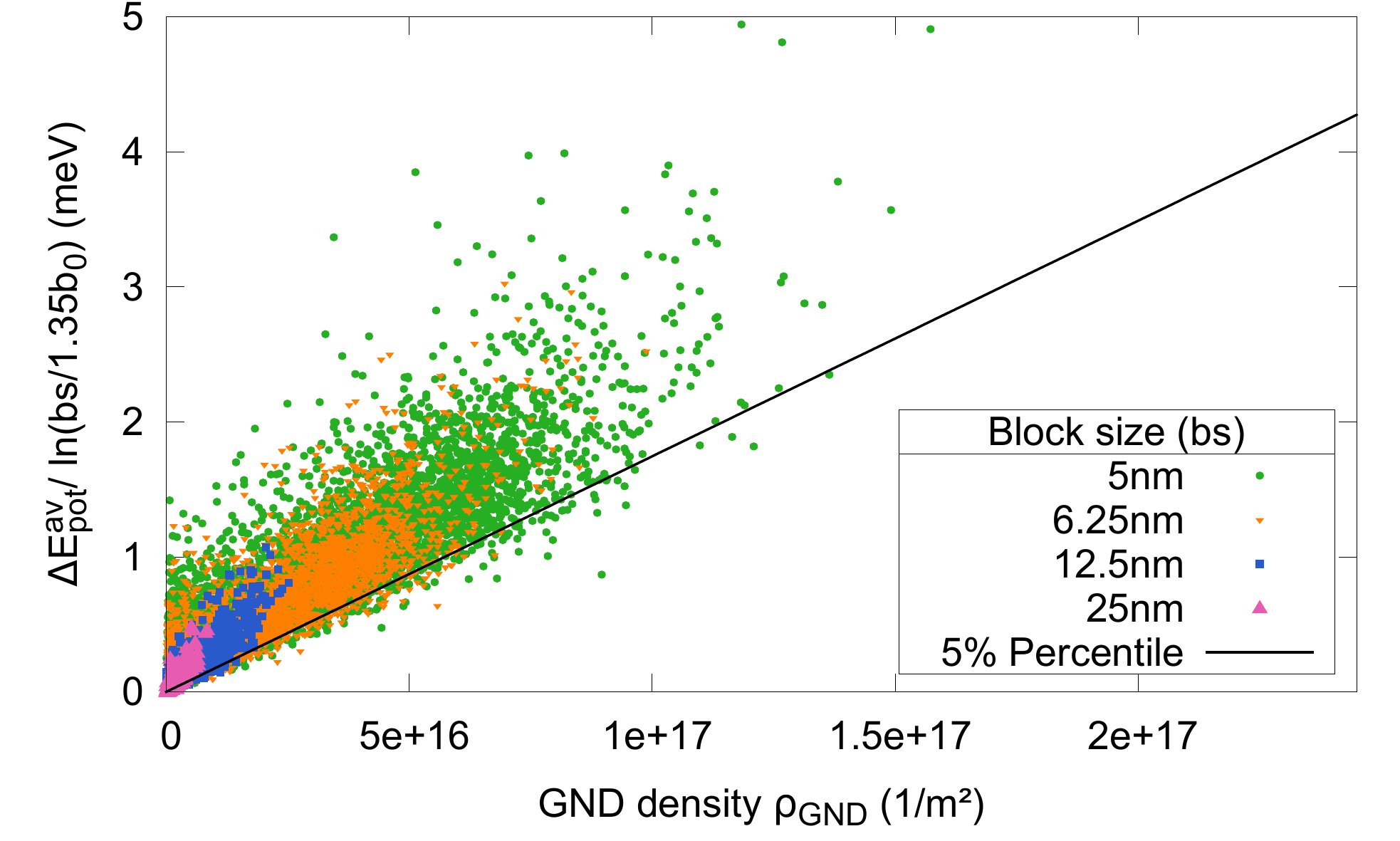}}\\
		\sidesubfloat[]{\label{fig:GND_rho_energy_subvolumesE}\includegraphics[width=0.46\textwidth]{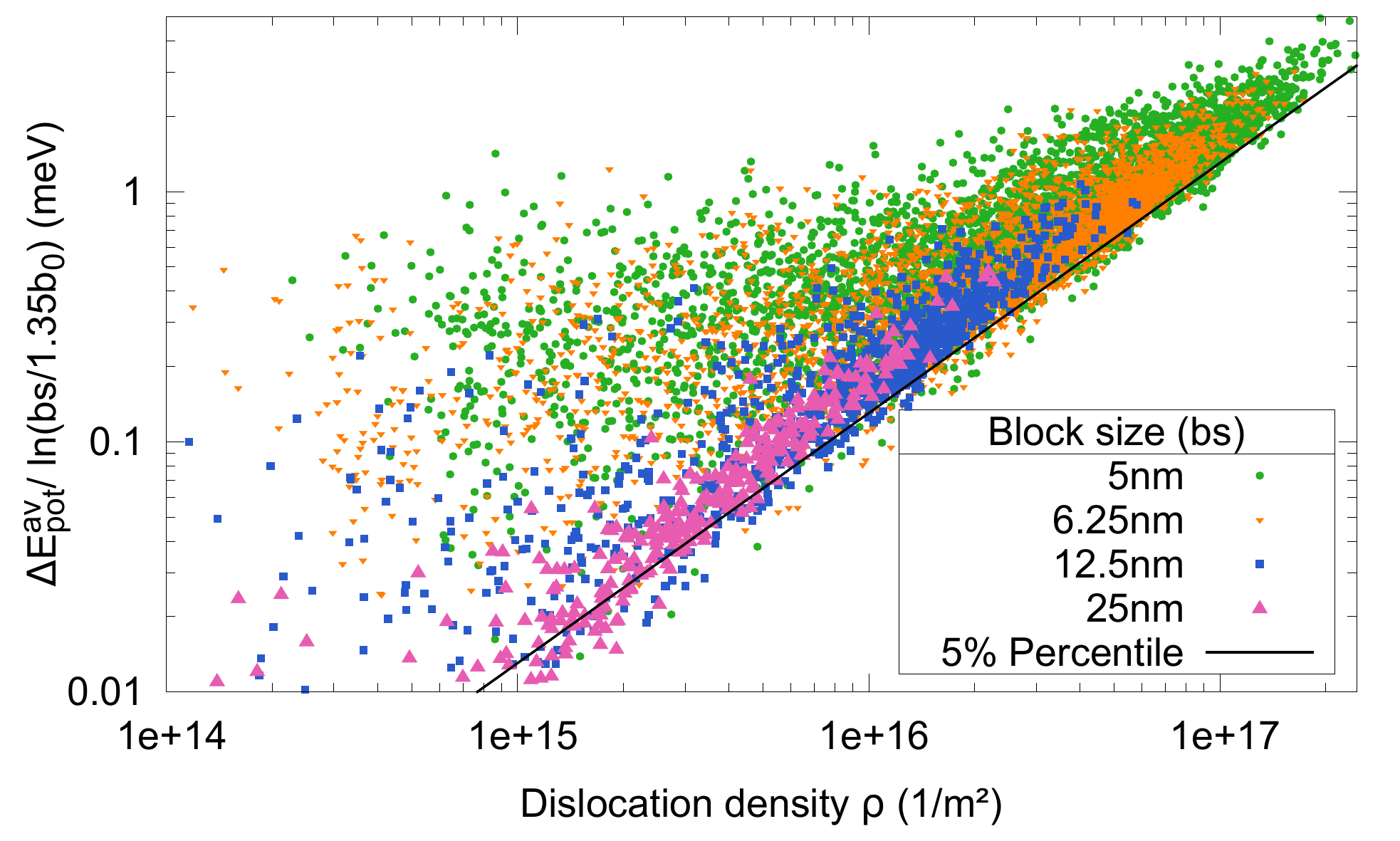}}
		\sidesubfloat[]{\label{fig:GND_rho_energy_subvolumesF}\includegraphics[width=0.46\textwidth]{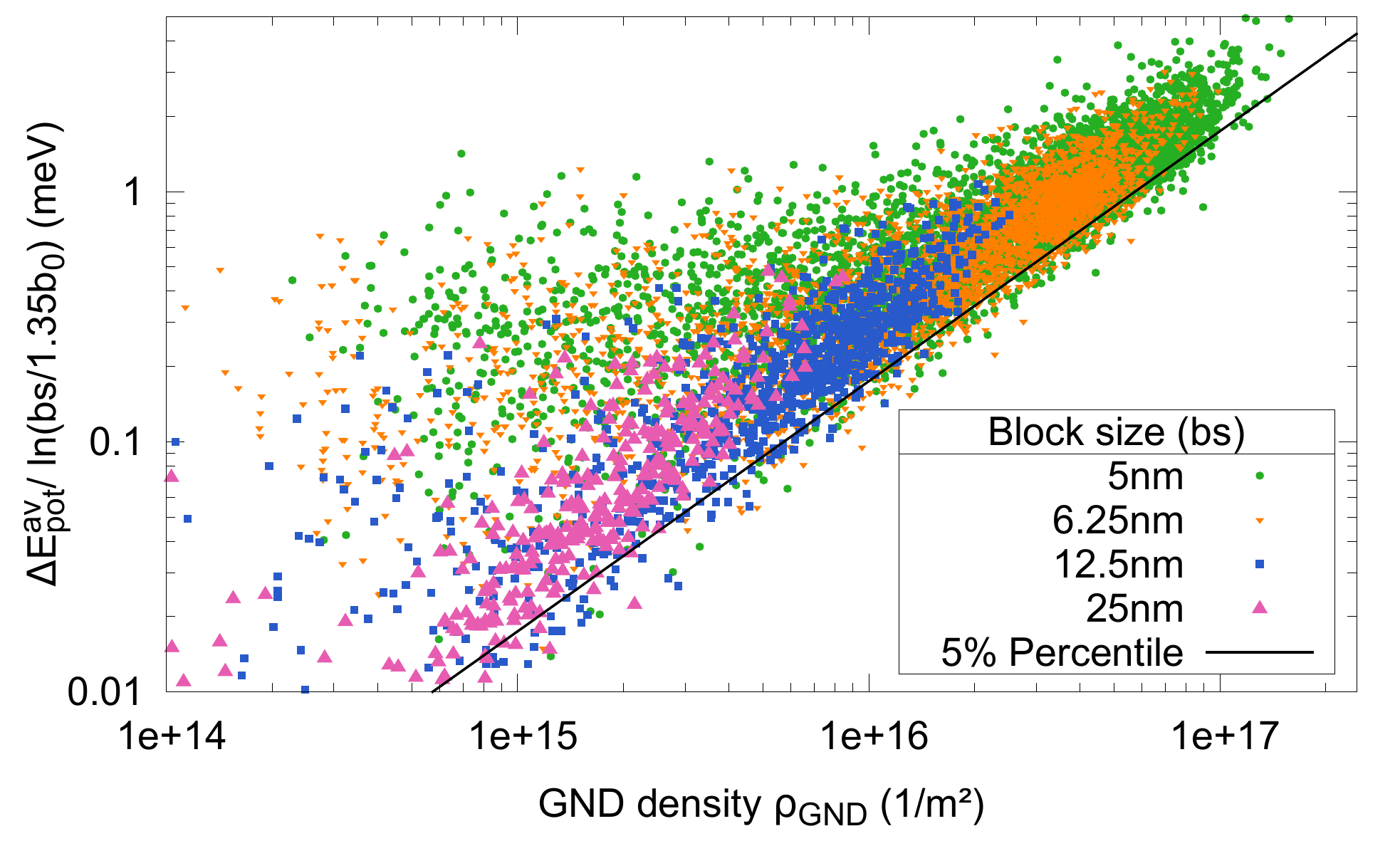}}\\
	\caption{Dislocation densities $\rho$ and $\rho_{\mathrm{GND}}$ in subvolumes: \protect\subref{fig:GND_rho_energy_subvolumesA}+\protect\subref{fig:GND_rho_energy_subvolumesB} linear scaling of the densities to the free energy per atom. In \protect\subref{fig:GND_rho_energy_subvolumesC}+\protect\subref{fig:GND_rho_energy_subvolumesD} a logarithmic correction factor based on the block size is applied. The same values are shown in a logarithmic scale in \protect\subref{fig:GND_rho_energy_subvolumesE}+\protect\subref{fig:GND_rho_energy_subvolumesF}.}
	\label{fig:GND_rho_energy_subvolumes}
\end{figure}
In Figs.~\ref{fig:GND_rho_energy_subvolumesA} and~\ref{fig:GND_rho_energy_subvolumesB} the relationship between the free energy and the densities $\rho$ and $\rho_{\mathrm{GND}}$ is plotted.
Due to the use of different block sizes, a wide range of densities is observed that varies over more than three orders of magnitude of $10^{14}$\,\si{\meter\tothe{-2}} to $10^{17}$\,\si{\meter\tothe{-2}}.
Several observations are made on these results.
As expected, smaller blocks contain in tendency very high dislocation densities where often $\rho \approx \rho_{\mathrm{GND}}$.
Furthermore these associated energies scatter significantly, but the scatter decreases noticeably for large blocks.
In small blocks, any defect located just outside of this volume can bias the measured energy severely without being observable in the dislocation densities and thermal fluctuations are becoming more pronounced as the number of atoms in the volume decreases.
In contrast, larger volumes are more independent of these deviations due to the larger number of atoms of which the average energy is measured.
Nonetheless, in both plots a lower limit for the free energy that scales linearly with the dislocation density is clearly observed.
This lower limit is approximated in the plots by fitting a linear function $f(x) = ax$ as the 5\% percentile, thus 95\% of the measured values are larger than the estimation by this function.

Using this approximation of a lower limit, a size effect is observed.
Whereas the free energy to dislocation density of large blocks (\SI{12.5}{\nano\meter} and \SI{25}{\nano\meter}) and small blocks (\SI{6.25}{\nano\meter} and \SI{5}{\nano\meter}) exhibit both a linear trend, they differ significantly in the slope.
Similar to the case of the full domains and in respect to Eq.~\ref{Eq:ElasticEnergy}, a logarithmic correction factor has been introduced in Figs.~\ref{fig:GND_rho_energy_subvolumesC} and~\ref{fig:GND_rho_energy_subvolumesD} with the factor $R$ defined in this case as being equal to the individual block size.
The same data using a logarithmic scale is shown in Figs.~\ref{fig:GND_rho_energy_subvolumesE} and~\ref{fig:GND_rho_energy_subvolumesF}.
With this modification, a clearly linear dependency between the dislocation density $\rho$ and the free energy is observed that covers all measured block sizes and the full range of dislocation densities $\rho$.
Especially the blocks of \SI{25}{\nano\meter} are in excellent agreement to the approximated lower limit of all measured values.
In contrast to the clear linear relationship between energy and $\rho$, the case is less clear $\rho_{\mathrm{GND}}$.
The deviations to the linear fit are more obvious and can be explained in the size dependency of $\rho_{\mathrm{GND}}$.
For smaller blocks $\rho_{\mathrm{GND}}$ tends to be equal to $\rho$, whereas the values deviate for larger blocks, e.g.\ the average for the \SI{12.5}{\nano\meter} is $\rho \approx 3\rho_{\mathrm{GND}}$, and thus $\rho$ does not scale linearly with $\rho_{\mathrm{GND}}$ across all scales.

In summary, the free energy is observed to scale linearly with the dislocation density $\rho$ across all length scales once a volume dependent factor is taken into account.
This observation holds true for cases with highly inhomogeneous distributions of dislocations in the full domains as well as in medium sized domains with rather homogeneous dislocation densities and even for very small domains containing typically only single dislocations. 


\section{Concluding remarks}
\label{Sec:Conclusion}
Very large scale atomistic simulations combined with detailed analysis methods have been used to study the relationship between dislocation microstructures and the associated free energy.
Since no \textit{a-priori} assumptions are made on this relation, it can be assumed as an \textit{ab-initio} approach for plasticity.

The results obtained from both the full sample sizes as well as from smaller subvolumes strongly indicate a linear relation between the total dislocation density $\rho$ of self-assembled dislocation networks and the free energy.
Furthermore, a logarithmic size dependent factor for an internal length scale $R$ is required to estimate the energy stored in elastic strains accurately.
Thus an energy function of the dislocation density in the form of Eq.~\ref{Eq:energyDensPsi} is here proposed as
\be
\psi = \psi(\rho) = A \cdot G b^2 \rho\cdot \ln(R/(r_0)) \quad,
\ee
where $A$ is a material dependent constant.
In the ten full sized samples the averaged measured value is $A=0.103$ with a standard deviation of $\sigma=0.0085$, where $\psi$ includes the energy of dislocation cores, stacking faults and the elastic energy.
Our MD simulations do not support a specific contribution of GNDs such that the free energy of dislocations networks appears to be only a function of the total dislocation density.
It has been shown that this model is able to describe the free energy for highly varying volumes and dislocation densities.
Furthermore, the model appears valid, both for highly inhomogeneous distributions of dislocations observed for nanoindentation, where dislocations are primarily located in the plastic zone under the indenter tip and glide of prismatic loops, as well as for the far more homogeneous distributions typically observed in small volumes.
This simple model can be related immediately to the elastic energy of a single straight dislocation as given by Eq.~\ref{Eq:ElasticEnergy}, and thus the total free energy can be considered as a superposition of the contribution of the individual dislocations.
While the elastic energy is including a size dependent part, the dislocation core energy is observed to scale directly linearly with the dislocation density $\rho$.
However, since this value is typically small in comparison to the elastic energy, the effect can be omitted.
Similarly, the factor based on the internal length scale $R$ may be omitted in continuum modeling if only the gradients of free energies are considered as along as $R$ is assumed to be constant.

A quadratic relation of the GND density to the free energy, which has been suggested frequently in strain gradient plasticity models, is not supported by our observations.
It is even questionable if such a form is meaningful at all, since the ratio of dislocation density to GND density is strongly affected by the choice of a reference volume.
For sufficiently small volumes every dislocation is geometrically necessary, thus different contributions of GND and scalar dislocation densities to the free energy are not plausible as long as no further criteria, like for example a minimum volume, are defined.
In particular, the model proposed here is very similar to the one proposed by Berdichevsky \cite{Berdichevsky2006}, which assumes a linear function for $\rho$ as long as the value is far smaller than the saturation value of the dislocation density.
Notably, the saturation value in the model by Berdichevsky is explicitly included as a threshold at which deformation cannot be accommodated by dislocations alone and other defects like subgrains must develop -- a mechanism that is actually observed in the atomistic simulations.

The irrelevance of the GND density for an energy function has several implications for the application in crystal plasticity.
The dislocation density $\rho$ is defined independently of a slip system that is otherwise needed to derive a proper norm of the tensorial dislocation density or the contribution of individual slip systems.
However its evolution must be modeled appropriately in order to estimate the statistically stored dislocations, which in contrast to the GND density cannot be derived from the deformation of a crystal.

In conclusion, the increasing computational resources provide the necessary tools to reach towards length scales in atomistic simulations on the order of \SI{100}{\nano\meter} that can be related to continuum scales.
In combination with the progress that has recently been made in analysis methods to quantify deformation and to identify dislocation networks in such simulations, it has become possible to measure related quantities directly which are otherwise unobtainable experimentally and are accurate enough to information that is required for scale bridging modeling.

\section{Acknowledgements}
This work is financially supported by ThyssenKrupp Steel Europe AG.
Part of the simulations was performed on JUQUEEN supercomputer facility at Forschungszentrum J\"ulich and provision of about 320,000 core-hours compute time is highly acknowledged by the authors.

\bibliography{literature}
\end{document}